\documentclass[aip, amsmath,amssymb, preprint]{revtex4-1}

\usepackage{CJK}
\usepackage[final]{graphicx}
\usepackage{hyperref}
\usepackage{color}
\usepackage{enumerate}
\usepackage{xr}
\usepackage{dcolumn}
\usepackage{bm}

\usepackage[utf8]{inputenc}
\usepackage[T1]{fontenc}
\usepackage{mathptmx}
\usepackage{etoolbox}
\usepackage{hyperref}

\usepackage{graphicx}
\usepackage{microtype}

\makeatletter
\def\@email#1#2{%
 \endgroup
 \patchcmd{\titleblock@produce}
  {\frontmatter@RRAPformat}
  {\frontmatter@RRAPformat{\produce@RRAP{*#1\href{mailto:#2}{#2}}}\frontmatter@RRAPformat}
  {}{}
}%
\makeatother

\begin{document}

\begin{CJK*}{GB}{} 

\preprint{STS/Pioneer-OpticalDesign}

\title[]{Optical design for the single crystal neutron diffractometer Pioneer}

\author{Yaohua Liu*}
\email[Author to whom correspondence should be addressed: ]{liuyh@ornl.gov}
\affiliation{Second Target Station, Oak Ridge National Laboratory, Oak Ridge, Tennessee 37831, USA}%

\author{Peter Torres}
\affiliation{Second Target Station, Oak Ridge National Laboratory, Oak Ridge, Tennessee 37831, USA}%

\author{Scott Dixon}
\affiliation{Second Target Station, Oak Ridge National Laboratory, Oak Ridge, Tennessee 37831, USA}%

\author{Cameron Hart} 
\affiliation{Second Target Station, Oak Ridge National Laboratory, Oak Ridge, Tennessee 37831, USA}%

\author{Darian Kent}
\affiliation{Second Target Station, Oak Ridge National Laboratory, Oak Ridge, Tennessee 37831, USA}%

\author{Anton Khaplanov} 
\affiliation{Second Target Station, Oak Ridge National Laboratory, Oak Ridge, Tennessee 37831, USA}%

\author{Bill M$^\textrm{c}$Hargue} 
\affiliation{Second Target Station, Oak Ridge National Laboratory, Oak Ridge, Tennessee 37831, USA}%

\author{Kumar Mohindroo} 
\affiliation{Second Target Station, Oak Ridge National Laboratory, Oak Ridge, Tennessee 37831, USA}%

\author{Rudolf Thermer} 
\affiliation{Second Target Station, Oak Ridge National Laboratory, Oak Ridge, Tennessee 37831, USA}%

\date{\today}

\begin{abstract}
Pioneer is a single-crystal neutron diffractometer optimized for small-volume samples and weak signals at the Second Target Station (STS) at Oak Ridge National Laboratory. This paper presents the preliminary optical design progress, focusing on the rationale behind key design choices. It covers the T$_0$ and bandwidth disk choppers, guide and beam control system, incident-beam polarizer, scattering beam collimators, and additional strategies. The chopper locations are selected to maximize neutron transport while taking advantage of standardized shielding structures. To accommodate the maintenance shield, shutter, and polarizing V-cavity, the guide design includes significant gaps. When these optical components are moved out of the beam path, oversized collimators, rather than guides, will be translated in. Pioneer will utilize slit packages to control beam size and divergence, and a translatable polarizing V-cavity. Absorbing panels are strategically placed near the end station to minimize background. An oscillating radial collimator, operating in a shift mode, will be used with the vertical cylindrical detector, while a fixed multi-cone collimator will be used with the bottom flat detector. These collimators will enable the detection of weak signals when complex sample environments are used.

\end{abstract}

\maketitle
\end{CJK*}

\section{introduction}
Pioneer is a time-of-flight single-crystal neutron diffractometer designed for the Second Target Station (STS) of the Spallation Neutron Source. Neutron scattering is a unique tool for materials characterization. Neutrons are non-destructive, penetrate deeply into matter, and are highly sensitive to light elements and magnetism~\cite{carpenter2015elements, adams2020first}. Leveraging the STS's high-brilliance source~\cite{STSConcept2020}, Pioneer will be capable of investigating the chemical and magnetic structures of X-ray-sized single crystals (0.001~mm$^3$) and small-area ($\sim5\times 5$~mm$^2$) epitaxial films as thin as 10~nm. This unprecedented capability overcomes the longstanding barrier of large sample volumes required for neutron scattering experiments, enabling researchers to tackle scientific challenges across diverse fields, from energy and quantum materials at an earlier stage to the characterization of thin films and heterostructures~\cite{liu2022Pioneer}. Pioneer will also facilitate experiments with complex sample environments and on tiny mineral inclusions formed naturally under extreme conditions that are difficult to replicate in laboratories. The instrument design is aimed at optimizing signal-to-noise ratio, minimizing systematic errors, enhancing operation reliability, simplifying implementation, and reducing costs while meeting engineering constraints and safety standards.

To match its broad science cases, Pioneer will use multiple slit packages and apertures to adjust the beam size and divergence, and provide a polarized incident beam option. The instrument requires a high flux on samples to enhance signals, a low flux outside the spatial region of interest (ROI) to minimize background, and a uniform beam profile within the required phase-space to reduce experimental artifacts and systematic errors. Our previous study on guide options~\cite{liu2024general} demonstrated that a straight beamline with a modified elliptical guide, despite requiring a T$_0$ chopper to block high-energy neutrons and gamma rays~\cite{carpenter2015elements}, is highly effective for transporting both thermal and cold neutrons from the compact STS moderator. Pioneer's guide system will adapt this option, consisting of two half-elliptical guides connected by multiple tapered sections containing gaps to accommodate necessary optical components. To ensure long-term operational reliability, we have strategically avoided guides at locations of the maintenance shield, shutter, and polarizer. When these components are out of the beam path, oversized collimators, rather than neutron guides, will translate into place. These collimators act as apertures to eliminate undesired neutrons/gamma rays reaching the end station. We have optimized the guide geometry to mitigate the effects of the large in-guide gaps~\cite{liu2025incident}. Additionally, Pioneer will use scattering beam collimators to enable measurements of weak scattering signals from samples using versatile sample environments. These design strategies can be applied to other neutron scattering instruments to enhance the sensitivity and accuracy in measuring weak signals using bulky sample environments.

Pioneer's optics design has evolved through multiple iterations. This paper outlines the design progress to date and the underlying reasoning for our choices. Section~\ref{sec:layout} introduces the instrument layout and key optical components. Section~\ref{sec:chopper} details the placement of the bandwidth-limiting and T$_0$ choppers. Section~\ref{sec:beamcontrol} describes the beam transport and control system. Section~\ref{sec:polarizer} discusses on the incident beam polarization system. Section~\ref{sec:endstation} focuses on the components located near the end station that are used to reduce background, including shielding and scattering beam collimators. Additional considerations are presented in Section~\ref{sec:discussion}.

\section{Instrument specification and layout}{\label{sec:layout}}
\begin{figure*}
\includegraphics[width=0.85\textwidth]{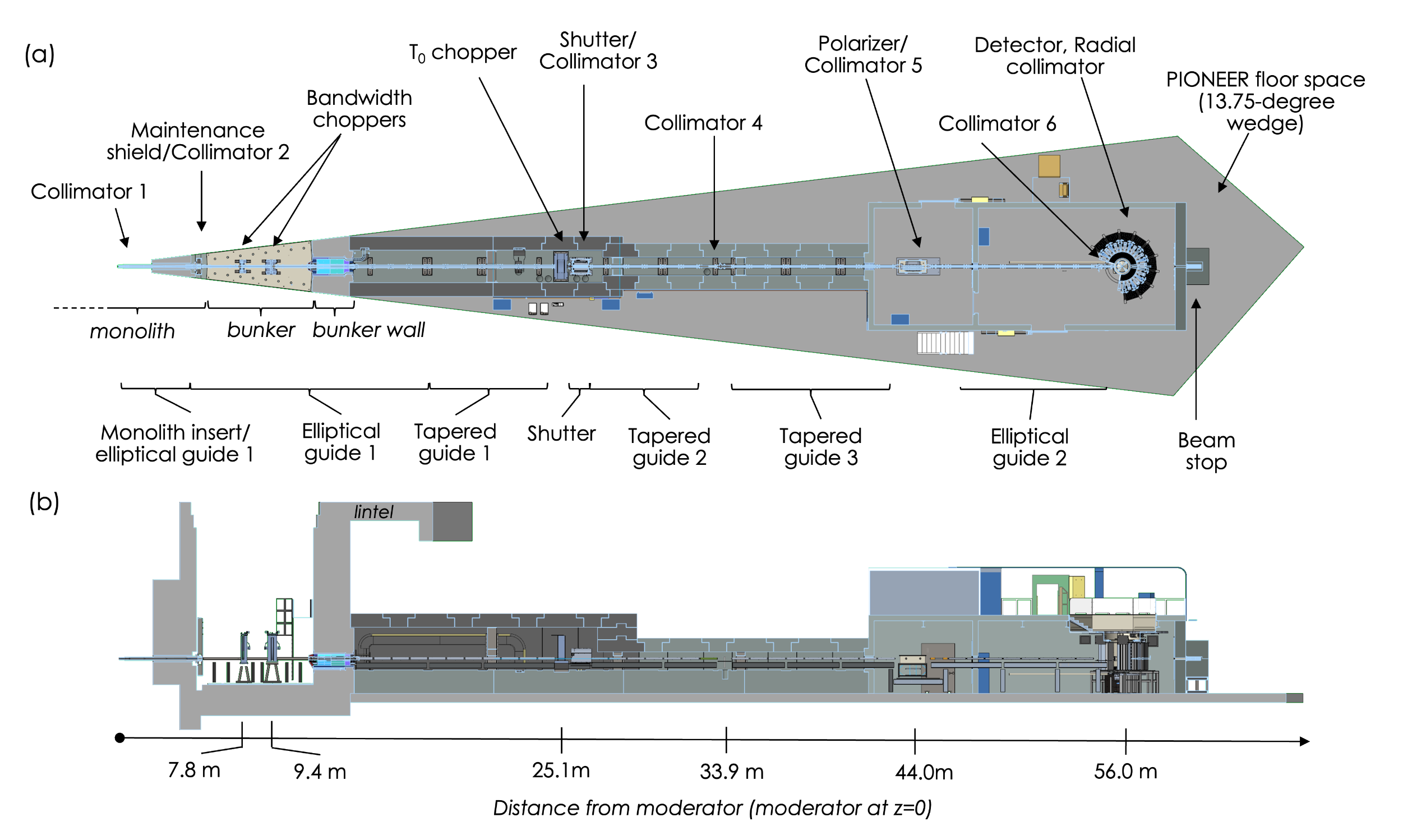}
\caption{\label{fig:layout} (a) The plan view and (b) the side view of the Pioneer instrument layout. The STS cylindrical moderator locates at $z=0.0$~m, and the nominal sample location is at $z=56.0$~m. The locations of key optical components are labeled, including the maintenance shield, the bandwidth-limiting and T$_0$ choppers, the shutter, the incident-beam polarizer, the cylindrical detector, the radial collimator and the beam stop. The slit locations are not indicated here for clarity but are shown in Fig.~\ref{fig:slitloc}. The different colors of the guide shielding and the cave wall denote the regular concrete (light gray) and high-density concrete (dark gray) used.}
\end{figure*}

\begin{table}[t]
\caption{\label{tab:techspec}Pioneer's specification. See the text for details.}
\begin{tabular}{l|c|l}
\hline \hline
parameter             & value               & comment\\
\hline
moderator             & parahydrogen      & cylindrical\\
$\lambda$ range       & 1.0 - 6.0~\AA     & \\
$\lambda$ bandwidth   & 4.3~\AA           & nominal 4.6~\AA \\
$\delta \lambda / \lambda$ &  $< 0.3\%$   &   \\ 
\hline
$L_1$                  & 56.0~m            & moderator to sample \\
$L_2$                  & 0.8 ~m            & sample to detector\\ 
\hline
detector coverage     &                   & with collimators  \\ 
\hspace{10pt}side cylindrical &  5.9~sr   &  \\ 
\hspace{10pt}bottom flat      &  0.7~sr   &  \\ 
\hline
beam size           
                      & 1 - 5 mm          & uniform region\\
\hline
$Q$ coverage          && \\
\hspace{10pt} nonpolarized beam &  0.1 - 12.5~\AA$^{-1}$  &   \\
\hspace{10pt} polarized beam    &  0.1 - 10.0~\AA$^{-1}$  &  \\  
\hline                 
beam divergence       && FWHM \\
\hspace{10pt} high-flux         &    0.7$^\circ$          & HF mode\\
\hspace{10pt} high-resolution   & $< 0.3^\circ$           & HR mode \\
\hline
Flux               &  1.2e9 n$\cdot$cm$^{-2}\cdot$s$^{-1}$ & HF mode\\  
\hline \hline
\end{tabular}
\end{table}

The key specifications of Pioneer are summarized in Tab.~\ref{tab:techspec}. Our decision to use the STS cylindrical moderator is driven by the required wavelength resolution while maintaining a broad wavelength band. The moderator-to-sample distance ($L_1+L_2$) is 56.80~m, giving rise to  $\delta \lambda / \lambda < 0.3\%$~across the entire operational wavelength range (see Fig.~\ref{fig:TOFRes} in the supplementary materials). Such a long beamline offers flexibility in guide design to reduce ambient background near the detector. Benefiting from the low source frequency, Pioneer will provide a wide wavelength band of 4.3~\AA~within a minimal range from 1.0 to 6.0~\AA. The 1.0-6.0~\AA~range is the minimum required to access the desired momentum transfer $Q$ coverage of 0.1-12.5 \AA$^{-1}$, as determined by the detector geometry. Our strategy therefore prioritizes beam transport performance within this range, although the guide system is capable of transporting neutrons over a broader wavelength range. The instrument features a large detector coverage with a nominal sample-to-detector distance of 0.8~m to accommodate versatile sample environments. The instrument supports tunable beam sizes and divergences, and provides a polarized incident-beam option. The flux on a sample in the high-flux mode is $1.2\times 10^{9}$ n$\cdot$cm$^{-2} \cdot$ s$^{-1}$, estimated from Monte Carlo ray tracing simulations on the ideal beamline before considering guide misalignment and windows along the guide~\cite{liu2025incident}.

Figure~\ref{fig:layout} shows the instrument layout from the optical point of view. The moderator is located at the origin, and the nominal sample location is 56.00~m away. Desired neutrons will be transported through the guide system with multiple gaps in between to accommodate the maintenance shield, choppers, shutter, slits, monitors, and polarizer. The instrument will utilize two half-elliptical guide sections (EG1 and EG2) connected by three tapered guide sections (TG1, TG2 and TG3). The guide geometry was numerically optimized via Monte Carlo ray tracing simulations using McStas~\cite{willendrup2014mcstas} and MCViNE~\cite{lin2016mcvine}. The optimized guide starts inside the monolith insert at 2.60~m and ends at 54.45~m, i.e., 1.55~m from the nominal sample position, and more information can be found elsewhere~\cite{liu2025incident}. There are a polarizer cave and a detector cave at the end of the instrument. 

The monolith houses the target wheel, a rotating assembly of tungsten segments bombarded by protons to generate neutrons via spallation. The beamline specific optics starts with a monolith insert that will be filled with Helium gas and house one optics module. In addition to the first portion of the guide system, the optics module contains a carbon steel collimator starting (C1) at 0.95~m from the moderator to block undesired neutrons and gammas and to avoid excessive energy deposition on the substrate face of the guide. 

The maintenance shield is positioned right after the monolith wall, a standardized design for all STS instruments. Its primary function is to shield personnel performing maintenance activities inside the bunker from gamma radiation emitted from the target area when the proton beam is off. The bunker is a heavily shielded room surrounding the target monolith. It houses various optical components, allowing personnel access for maintenance without disassembling large amounts of shielding~\cite{STSConcept2020}. During normal operation, the gamma blocker of the maintenance shield will be lifted, and a carbon steel collimator (C2) will translate into the beam path, creating a 0.33~m gap without a guide.

Pioneer will install two bandwidth-limiting choppers inside the bunker: one single-disk large chopper (C1-SDL, or SD1 for short) at 7.78~m and one double-disk large chopper (C2-DDL, or DD2 for short) at 9.38~m. "Large" follows the STS standard naming scheme to distinguish chopper sizes~\cite{STSICS2024}. The locations denote the center of the choppers. There will be retractable beam monitors (M1, M2) immediately before SD1 and after DD2, respectively. The chopper and monitor combinations require in-guide gaps of 0.24~m and 0.27~m around the SD1 and DD2, respectively. 

The guide shielding starts immediately outside the bunker wall at 13.90~m and ends before the cave wall at 42.10~m. Key optical components inside the guide shielding include the T$_0$ chopper (C3-T0L, or T$_0$ for short) at 25.10~m, the shutter at 26.31~m, and two slit packages (S1, S2) at 25.70~m and 34.00~m, respectively. S1 is placed between the T$_0$ chopper and the shutter, and S2 is at the overlapped focal points of EG1 and EG2. During experiments, the shutter is opened and a 1.00-m long carbon steel collimator (C3) will translate into the beam path, which will reduce the high-energy neutrons and the gamma rays, either leaked or generated from the T$_0$ chopper. Another 1.00-m long carbon steel collimator (C4) is permanently installed before S2, and a retractable monitor (M3) is placed after S2. 

The polarization system will start inside the polarizer cave. The polarization system comprises of a translatable, multi-channel polarizing supermirror V-cavity, an adiabatic radio-frequency (RF) spin-flipper, and guide-field magnets~\cite{mezei1989very, fitzsimmons2005application}. The 1.0-meter long V-cavity is housed within a vacuum vessel, connecting to the guide system with two gate valves. There is a retractable monitor (M4) immediately after the upstream gate valve. The assembly of the M4, gate valves, and polarization vessel requires an in-guide gap of 1.89~m. To use the non-polarized mode, the V-cavity will move out, and a 1.0-m long absorbing flight tube (C5) will translate into the beam path. Molded ZHIP panels are installed on the rear wall of the polarizer cave to absorb neutrons leaked from the polarizer and the guide. ZHIP stands for zero hydrogen in product and is a B$_4$C based composite material with a non-hydrogenous binder~\cite{ZHIP, abernathy2012design}.

A slit (S3) is installed after the second elliptical guide (EG2) at 54.80m, with a gate valve and a small chamber for a laser alignment system positioned between them. The gate valve isolates the guide system from the reentrant vessel during sample environment changes, while the laser alignment system provides an optical reference to indicate the incident neutron beam path. S3 is followed by a monitor (M5) and the final slit, S4, located at 55.37m. A permanent 0.44-m absorbing flight tube (C6) is installed between M5 and S4. The inner surfaces of C5 and C6 are coated with B$_4$C to eliminate unwanted stray neutrons.

The end station, centered at 56.0~m (nominal sample location), features a vacuum reentrant vessel to accommodate top-loading sample environments and an oscillating radial collimator (ORC) between the reentrant vessel and the vertical cylindrical detector. There is an insertable aperture within the vessel, which will have variable openings and aperture-to-sample distances. Pioneer also has a removable, flat detector beneath the sample position, which can be replaced with a roll-in goniometer system for high-throughput experiments using a six-axis robotic sample changer. The reentrant vessel will be removed to use the goniometer. A permanent beam monitor (M6) is placed inside the get-lost tube at the down-stream wall of the detector cave. 

\section{Choppers}{\label{sec:chopper}}

The STS moderator generates copious neutrons over a broad energy range from tens of $\mu$eV to tens of MeV~\cite{carpenter2015elements, STSConcept2020}. In contrast, neutron scattering instruments typically utilize only a narrow energy range for materials research, from several hundred $\mu$eV to a few eV~\cite{carpenter2015elements}. Fast neutrons and gamma rays produced by the high-energy proton beam colliding with the target are unsuitable for scattering experiments. However, they and their secondary particles can reach the end station and contribute to background through inelastic scattering interactions with the sample and its environment~\cite{jones1987HET, Santoro2022}. Therefore, these unwanted particles shall be eliminated from the beam path. In straight beamlines like Pioneer, a T$_{0}$ chopper is commonly used to block the initial burst of fast neutrons and gamma rays~\cite{carpenter2015elements}. To select neutrons with the desired wavelength range, bandwidth-limiting choppers are also employed, which are coated with thin-layer absorbing materials to block unwanted slow neutrons while allowing the desired ones to pass through~\cite{carpenter2015elements}. 

Pioneer will use a T$_0$ chopper and two disk choppers, SD1 and DD2. Ideally, all choppers would be located inside the bunker to minimize downstream shielding requirements. However, this positioning introduces in-guide gaps of 0.62~m, 0.07~m, and 0.13~m, respectively, which could adversely affect neutron transport performance. Our Monte Carlo ray tracing simulations indicate that neutron transport is more sensitive to larger gaps in the elliptical guide region than in the tapered region. The relatively small gaps for SD1 and DD2 can be mitigated by optimizing the elliptical guide parameters\cite{liu2025incident}. On the other hand, placing the T$_0$ chopper inside the bunker results in divergence gaps, reducing flux at the sample and introducing artifacts in single-crystal diffraction experiments. As a result, placing the bandwidth-limiting choppers inside the bunker while positioning the T$_0$ chopper outside is a better solution.

In addition to optimizing guide transport performance, we have also taken wavelength bands and building constraints into account to determine the optimal chopper locations, as outlined below.

\subsection{T$_0$ chopper}
\begin{figure}
\includegraphics[width=0.48\textwidth]{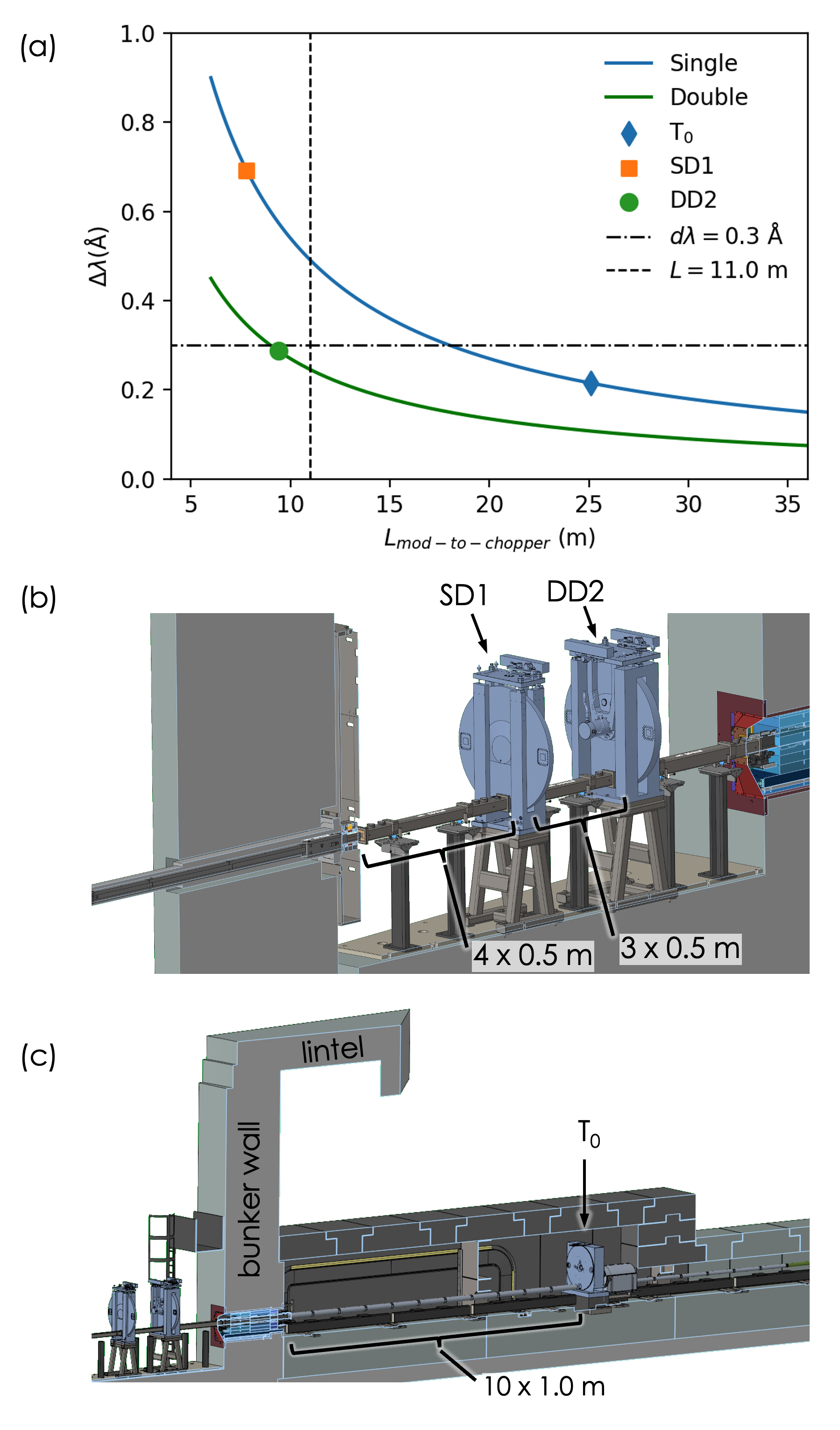}
\caption{\label{fig:Chopper_transition}(a) The transition width $\Delta\lambda$ as a function of the chopper-to-moderator distance, based on the STS conceptual large chopper design. `Single' refers to configurations with either a single T$_0$ chopper or a single-disk chopper, while `Double' indicates configurations with counter-rotating double-disk choppers. The symbols represent the selected chopper locations. The dashed vertical line marks the maximum allowable chopper position within the bunker, and the dash-dotted horizontal line represents the target maximum transition bandwidth of 0.3~\AA.  (b), (c) The selection of choppers has considered the monolith and bunker structures, as well as the number of supermirror elements.}
\end{figure}

A T$_0$ chopper is made of materials with high cross-sections for high-energy neutrons to block the prompt pulse, which also eliminates unwanted neutrons and gammas from the incident beam. The STS T$_{0}$ chopper features a 30-cm Inconel X-750 block~\cite{STSChopper2018}. Similar design variants have been used at different facilities~\cite{jones1987HET, itoh2012t0, violini2014investigation}. Per the STS chopper requirement, the edge of the chopper blade shall sweep through the entire beam within $\Delta t = 1.365$~ms while operating at 15~Hz, which is equal to the sweep time specification in the first target station at SNS~\cite{STSConcept2020, STSChopper2018}.The width of the partial transmission band during the chopper's sweep, $\Delta \lambda$, is given by, $\Delta \lambda = \frac{h \times \Delta t}{m L_{mod-to-chopper}}$, where $h$ is the Planck constant, $m$ is the neutron mass and $L_{mod-to-chopper}$ is the distance between the moderator to the T$_0$ chopper. 

Figure~\ref{fig:Chopper_transition}(a) shows $\Delta \lambda$ as a function of $L_{mod-to-chopper}$ for the STS chopper concept. The T$_0$ chopper is placed at 25.1~m from the moderator, allowing space for 10 standardized guide housings, each 1.0~m in length, between the bunker wall and the T$_0$ chopper (see Fig.~\ref{fig:Chopper_transition}(c)). Since Pioneer does not use epithermal or hot neutrons, we choose to use the T$_0$ chopper to block them, as disk choppers are generally less effective at attenuating these higher-energy neutrons\cite{jones1987HET}. To achieve the desired transition width of $\Delta \lambda \lesssim 0.3$~\AA, the T$_0$ chopper shall be positioned at least 17.9~m away from the moderator. At 25.1~m, $\Delta \lambda$ is 0.21~\AA, allowing the T$_0$ chopper to completely block neutrons up to below 0.3-0.4~\AA~but fully transmit neutrons above 0.6-0.7~\AA~for experiments. Positioning the chopper further upstream would obstruct easy access for installation and maintenance using the building crane. Moving it downstream necessitates a heavier Inconel block (to eliminate neutrons up to 0.3-0.4~\AA) and might increase the background level near the end station~\cite{wang2023physical}.

\subsection{bandwidth-limiting choppers}
\begin{figure}
\includegraphics[width=0.48\textwidth]{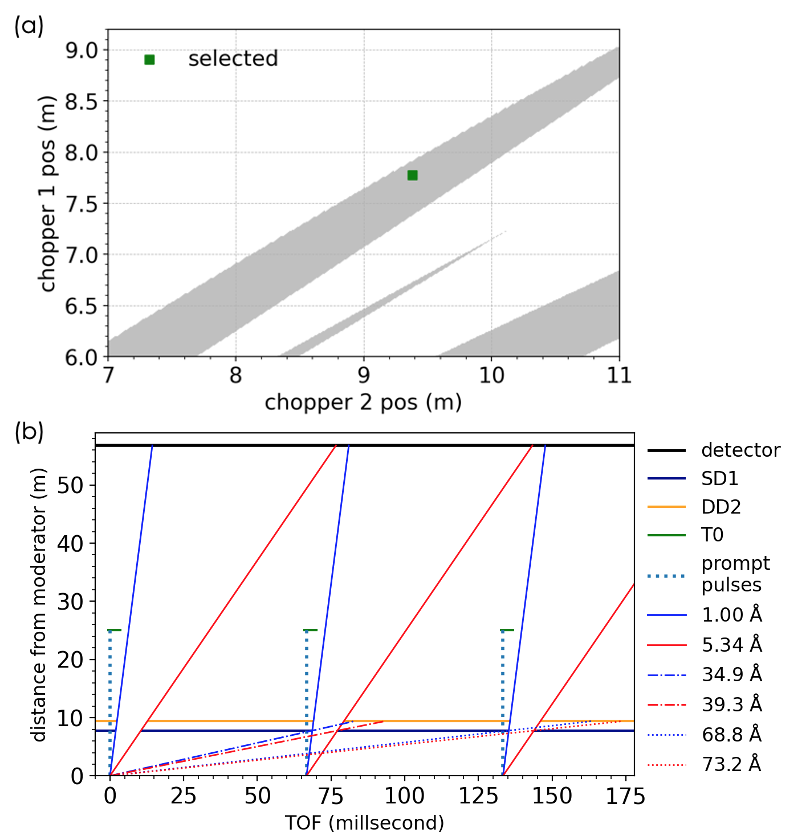}
\caption{\label{fig:Chopper_TOF}(a) The gray area indicates the working locations of two bandwidth-limiting disk choppers inside the bunker. The selected locations are labeled.  (b) Neutron time-schedule diagram showing the action of the T$_{0}$ and two disk choppers. STS will produce short pulses and the diagram has ignored the emission time. The trajectories of neutrons at the band boundaries from three consecutive pulses are shown by the blue and red lines.  The diagram also shows the long-wavelength neutrons leaked from SD1 but blocked by DD2. The T$_0$ chopper will fully block the fast neutron and partially block hot and epithermal neutrons up to 0.6-0.7~\AA.}
\end{figure}

The low-frequency STS source allows for flexible placement of the two bandwidth-limiting disk choppers. These choppers, which have wide openings, are used to define a broad range of neutron wavelengths or to prevent frame overlap~\cite{carpenter1984instrumentation}. Typically, strong, lightweight materials such as aluminum or carbon fiber are used to construct the rotors, while B10-epoxy serves as the neutron-absorbing coating to block unwanted slow neutrons~\cite{carpenter2015elements}.

The phase and angular opening of a disk chopper can be determined from the moderator-to-chopper distance $L_{mod-to-chopper}$, given the fixed moderator-to-detector distance, the source/chopper frequency, and the chosen wavelength band. Assuming the chopper running at the source frequency of 15~Hz and ignoring the emission time, the scheduled opening time of a chopper can be calculated. For example, as shown in Fig.~\ref{fig:Chopper_TOF}(b), SD1 transmits the desired band from 1.0 to 5.3~\AA, but it will also transmit long-wavelength neutrons around 37.0~\AA~and 71.0~\AA~from earlier pulses. Figure~\ref{fig:2BLchoppers} presents Monte Carlo ray tracing simulations, revealing significant leakage of long-wavelength neutrons when only one bandwidth-limiting chopper is used. The effective flux for time-of-flight diffraction~\cite{jauch1997prospects} includes a multiplication factor of $\lambda ^2$. If these cold neutrons were to reach the detector, Pioneer's detectors would record incorrect time-of-flights for these events, leading to a frame overlap issue~\cite{carpenter1984instrumentation}. Adding a second disk chopper blocks these leaked cold neutrons, with the location chosen based on its ability to block the leaked neutrons from the first chopper.

Our calculations accounted for potential leakage from the previous three pulses. Figure~\ref{fig:BLchop_loc} shows the results, suggesting that the first chopper shall be placed inside the bunker, while the second one can be positioned either inside or outside. To minimize downstream shielding requirements, we opted to place both disk choppers inside the bunker.

The transition width of the bandwidth-limiting choppers reduces the usable wavelength band for experiments, within which the beam is fully transmitted through the chopper. As shown in Fig.~\ref{fig:Chopper_transition}(a), placing the choppers inside the bunker results in a large transition width more than 0.5~\AA~for a single-disk chopper. However, this effect can be partially mitigated by using a double-disk chopper with two counter-rotating rotors, which reduces the transition width by half~\cite{STSChopper2018}. Only one double-disk chopper is needed to define the wavelength band, while the other can be an single-disk chopper to prevent frame overlap.  To achieve a sharper transition, the double-disk chopper should be positioned further downstream than the single-disk chopper. Figure~\ref{fig:DDchoice} compares simulated spectra for different chopper combinations, both using an single-disk chopper as the frame-overlap chopper. Replacing one single-disk chopper with a double-disk chopper increases the usable wavelength band from 4.0~\AA~to 4.3~\AA.

Using the choppers requires breaking the continuity of the guide system. However, the guide system is constructed from supermirrors, with each individual element having a maximum length of 0.50~m, as specified by manufacturers. Precisely aligning these supermirror elements is critical for optimizing guide performance, but this can be challenging. To a first-order approximation, flux reduction due to guide misalignment is proportional to the number of elements\cite{zendler2015generic}. Shorter elements increase the total number, exacerbating potential misalignment issues. Therefore, it is preferable to use as many 0.50~m elements as possible. Therefore, we have positioned the SD1 and DD2 choppers at 7.78~m an 9.38~m, respectively. At these locations, the number of 0.5~m elements is maximized inside the bunker, as shown in Fig.~\ref{fig:Chopper_transition}b. 

\section{Guide and beam control system}{\label{sec:beamcontrol}}
\begin{figure*}
\includegraphics[width=0.72\textwidth]{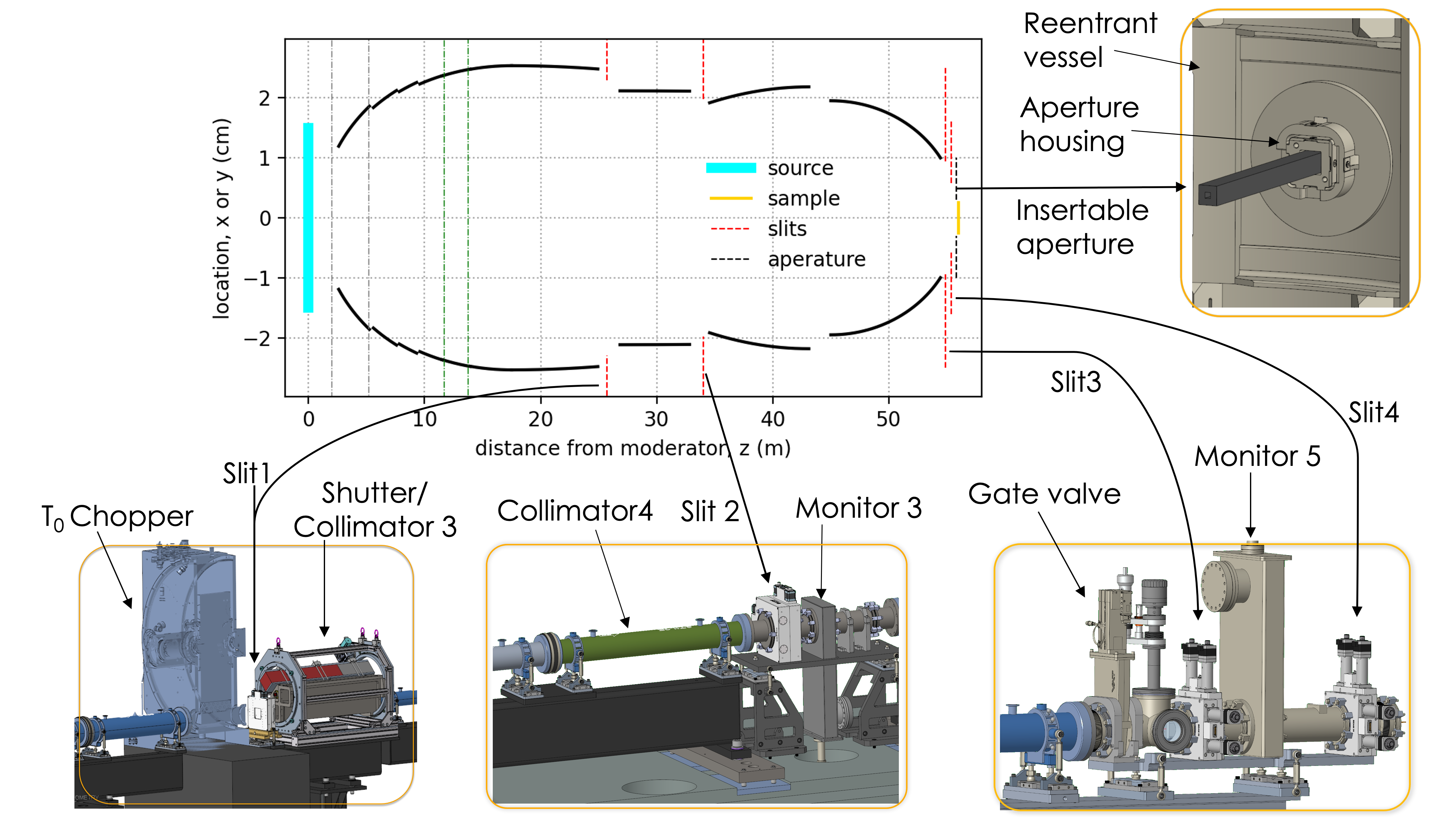}
\caption{\label{fig:slitloc}A schematic of the neutron guide system showing the positions of the four slit packages and the final insertable aperture. The figure also highlights the interfaces between the slit packages and the guide system, as well as the placement of the aperture within the reentrant vessel. The gray dot-dashed lines indicate the boundary of the monolith insert, while the green dot-dashed line marks the boundary of the bunker wall.}
\end{figure*}

\begin{figure*}
\includegraphics[width=0.96\textwidth]{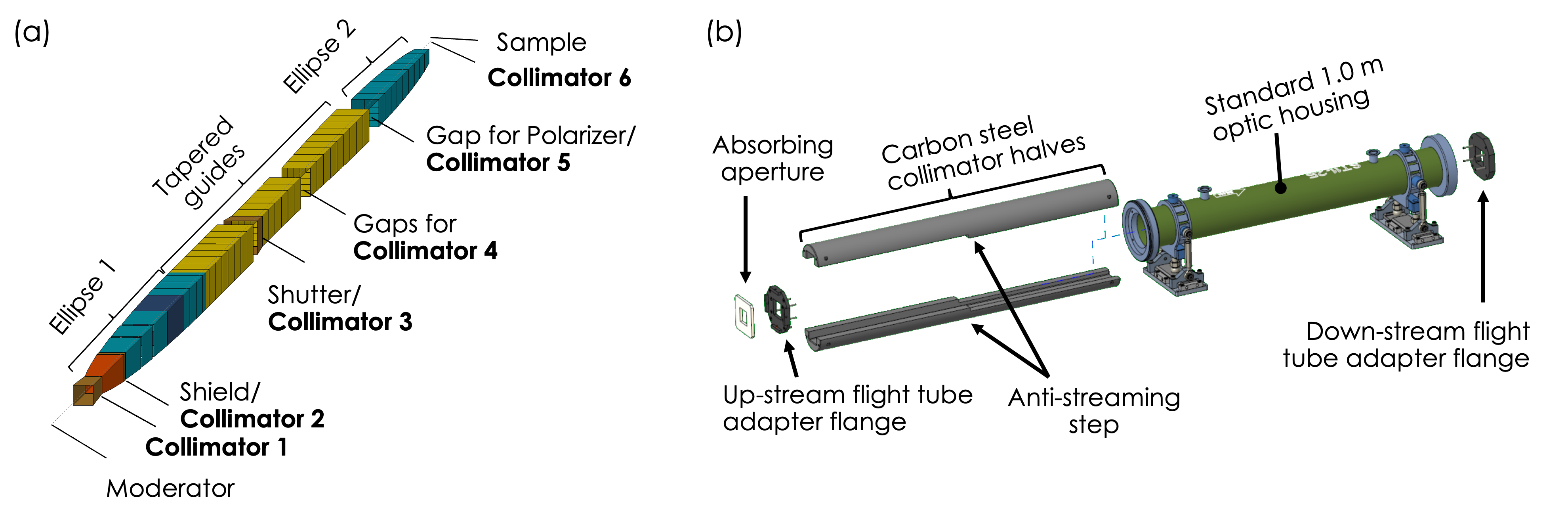}
\caption{\label{fig:inguidecol}(a) A 3D view of the guide system highlighting the locations of the in-guide collimators, with different guide sections shown in distinct colors. (b) Design of collimator C4 and its housing, featuring stepped collimator halves to block streaming paths.}
\end{figure*}

With neutron guides at a pulsed source, we can position the end station at a greater distance from the moderator, meeting the required wavelength resolution and reducing ambient background at the detector. The guide must be designed to transport the neutron beam while achieving the desired beam size, divergence, and homogeneity. Initial studies have shown that a straight beamline with elliptical guides can effectively meet Pioneer's requirements, which delivers high fluxes for both thermal and cold neutrons but requires a T$_0$ chopper. The relatively small cross-section of Pioneer's guide is related to the compact design of the STS high-brightness moderators~\cite{adams2020first} and the required phase space. With a total guide length of approximately 50~m, precise alignment during assembly and operation is critical~\cite{zendler2015generic,lin2023realistic}. To avoid the alignment challenges posed by translatable guides, Pioneer's guide will include large gaps to accommodate essential optical components, necessitating optimization of the guide geometry to compensate for these gaps. 

Pioneer's beam control system will provide tunable beam size and divergence to meet various experimental needs. We have chosen slit packages and apertures due to their simplicity and operational reliability. Pioneer will utilize commercially available, vacuum-compatible slit packages. These packages contain motorized, neutron-absorbing blades, allowing remote adjustment of the slit opening to control beam size and divergence. In addition, customized, exchangeable apertures with fixed openings and lengths matched to the sample environment will provide further beam profile tailoring. These apertures have standardized external dimensions for easy manual exchange within the aperture housing. Monte Carlo ray tracing simulations were performed to determine the optimal number and placement of the slit packages with the goal of delivering a high-flux beam to maximize the signal, filtering out unwanted neutrons early to reduce ambient background, and ensuring a uniform beam at the sample position to minimize systematic errors~\cite{liu2024general}. The positions and openings of the slits, along with the guide geometry, were optimized simultaneously. The system with four slit packages and a final insertable aperture was found to perform satisfactorily. Further details on the optimization methodology and results will be presented elsewhere~\cite{liu2025incident}.

Figure~\ref{fig:slitloc} shows the optimized guide geometry and slit locations. The guide features two half-elliptical sections sharing a focal point, connected by multiple tapered segments. Inside the bunker, the first half-elliptical guide is divided into three sections to accommodate the maintenance shield and the SD1 and DD2 choppers. These sections are short and of varying lengths, requiring customized housings for the guides, as shown in Fig.~\ref{fig:Chopper_transition}(b). Outside the bunker, the guide sections are longer, and we will use 1.00-m STS-standardized guide housings, each containing two 0.50-m supermirror elements. Figure~\ref{fig:slitloc} illustrates the integration between the slit packages and the guide system, showing seamless interfaces between the slits, apertures, and guide.

Pioneer will use six in-guide collimators, as indicated in Fig.~\ref{fig:inguidecol}(a). Two types of in-guide collimators will be used: high-Z material collimators (C1-C4) for shielding against high-energy neutrons and gamma rays, and guide-type collimators with absorptive surfaces (C5-C6) to filter out unwanted slow neutrons. For instance, Fig.~\ref{fig:inguidecol}(b) shows the design of C4, located between TG2 and TG3. C4 consists of a 1.0-m long carbon steel section housed within a standardized 1.0-m optical housing. Carbon steel is selected for C1-C4 components due to its comparable shielding performance to Cu and W~\cite{santoro2018study}, but lower cost. The collimator is divided into two carbon steel halves to block streaming paths. An absorbing aperture placed before C4 will further filter out unwanted slow neutrons.

\section{Incident Beam Polarization System}{\label{sec:polarizer}}
\begin{figure*}
\includegraphics[width=0.8\textwidth]{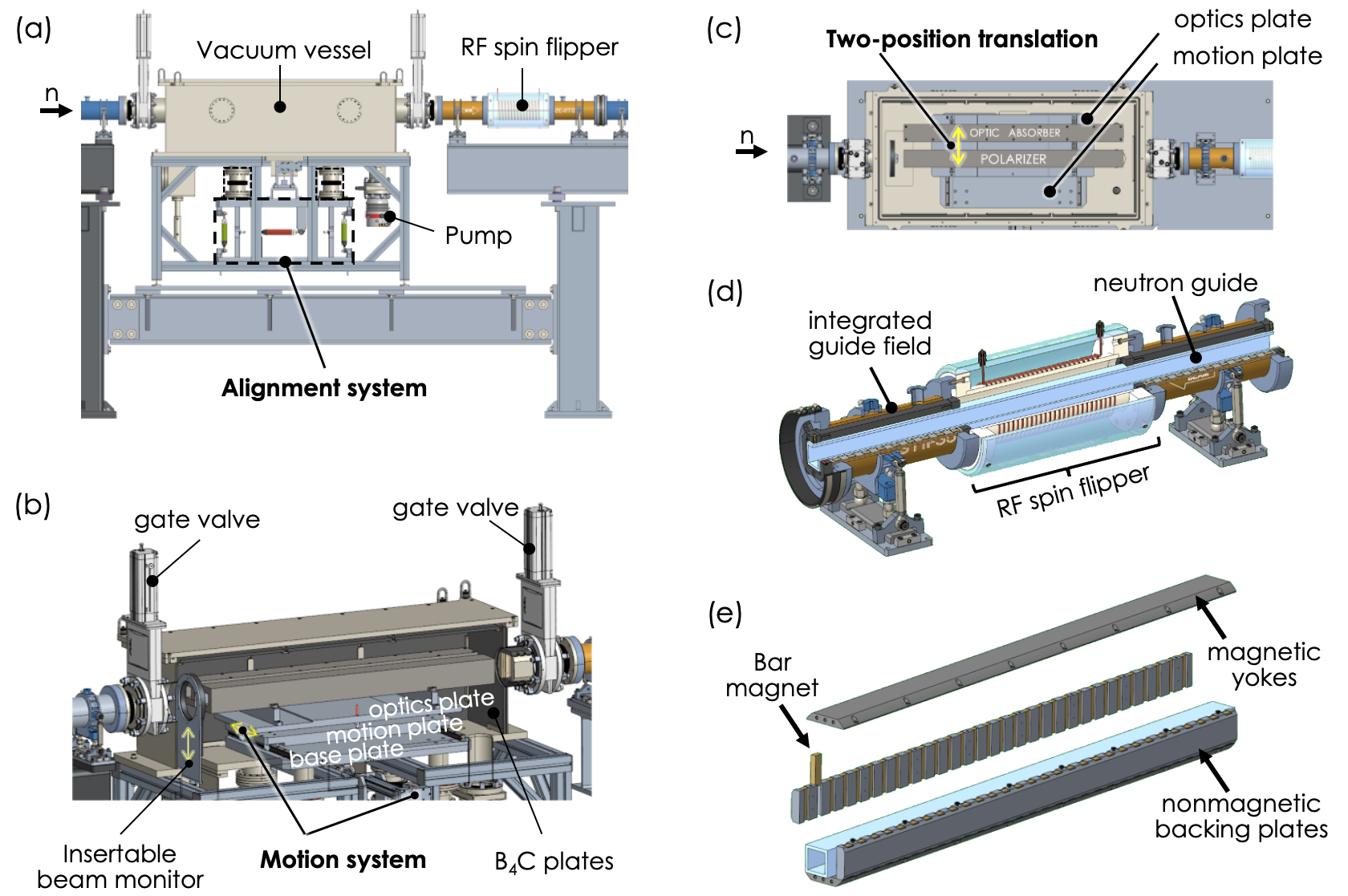}
\caption{\label{fig:polbox}(a) The incident beam polarization system, showing the locations of the polarizer vacuum vessel, RF spin-flipper, and their interfaces with the neutron guide system. (b) The translatable polarizing V-cavity housed within the vacuum vessel, connected to the guide system via two gate valves. The polarizer is mounted on an optics plate, which is kinematically coupled to a motion plate. (c) The motion system enables two-position translation, allowing either the V-cavity or the collimator to be positioned in the beam path. (d), (e) Conceptual designs of the spin-flipper and the guide field magnets.}
\end{figure*}

Figure~\ref{fig:polbox} shows the design of the Pioneer's polarization system, which comprises three key components: a translatable polarizing V-cavity, an adiabatic RF neutron resonant spin-flipper, and bar magnets that generate the necessary magnetic field along the neutron guide~\cite{fitzsimmons2005application}. Simulations show that the polarizing V-cavity outperforms a $^3$He filter over the broad wavelength band used at Pioneer~\cite{liu2025incident}.

Two primary factors determine the optimal placement of the polarizing V-cavity: beam divergence and available installation space. The V-cavity operates most effectively in regions of low beam divergence and requires substantial installation space. Therefore, the ideal location is within the tapered guide region, where beam divergence is minimal and guide performance is less sensitive to gaps. Pioneer will place the polarizer immediately before the second elliptical guide, minimizing the guide length requiring a magnetic field. As shown in Fig.~\ref{fig:polbox}(a), the RF adiabatic spin-flipper will be positioned at the start of the second elliptical guide. A dedicated cave will house the polarizer and the spin-flipper, facilitating easy access for maintenance and upgrades, as shown in Fig.~\ref{fig:layout}.

The entire polarizer assembly is housed within a vacuum vessel, which can accommodate a V-cavity up to 1.20~m and an insertable beam monitor. It connects to the neutron guide through two gate valves, providing a window-free optical path to minimize beam loss. This design enables polarizer maintenance without breaking the vacuum of the entire guide system, reducing operational downtime. The interior of the vacuum vessel is lined with reaction-bonded boron carbide (B$_4$C) plates to absorb scattered neutrons from the polarizer.

The polarizer alignment and motion system is critical to its performance. The system features six struts, providing six degrees of freedom to connect two parallel plates across the vacuum boundary: a base plate and a motion plate. This design isolates the alignment and motion systems from the vacuum vessel, ensuring stability. The polarizer is mounted on an optics plate that is kinematically coupled to the motion plate.  The system allows transverse translation relative to the neutron beam, enabling two positions: with either the polarizer or the absorbing collimator in the beam path, shown in Fig.~\ref{fig:polbox}(c). In non-polarized mode, the V-cavity is retracted, and the collimator is inserted into the beam path. Using an oversized collimator instead of a guide relaxes alignment constraints and reduces the risk of performance degradation while filtering out unwanted neutrons. The motion system provides positive position feedback, motion limits, and manual operation capability in case of motor failure, ensuring reliable operation under challenging conditions.

Figures~\ref{fig:polbox}(d) and (e) present the conceptual designs for the adiabatic RF spin-flipper and the guide field. The RF spin-flipper operates over a broad wavelength range~\cite{fitzsimmons2005application} and provides an open beam path for optimal neutron transport. The guide field will be generated by permanent magnets mounted on non-magnetic plates.

\section{End Station}{\label{sec:endstation}}
\begin{figure}
\includegraphics[width=0.48\textwidth]{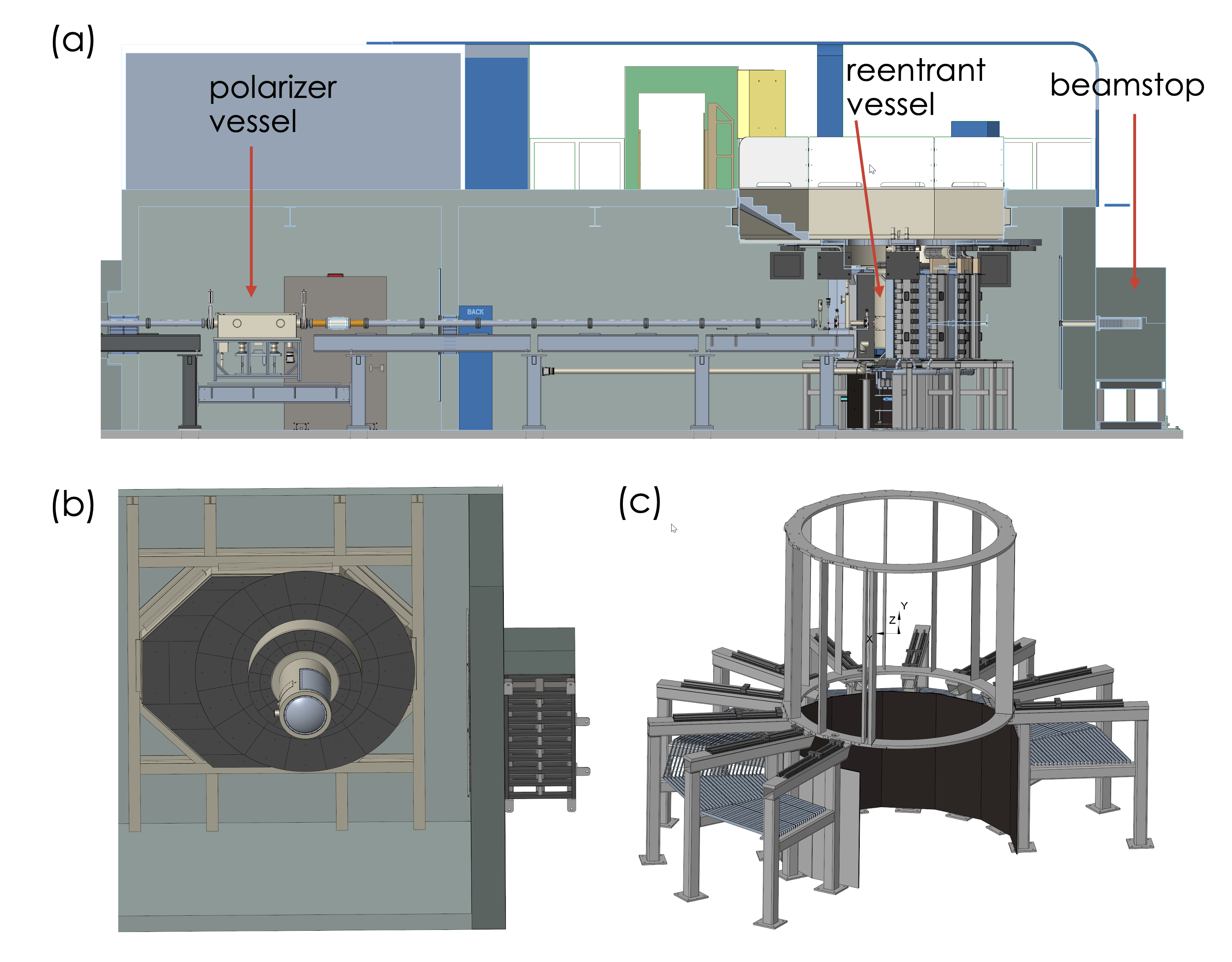}
\caption{\label{fig:B4C} Shielding components near the end station. (a) View of the polarizer and detector caves. ZHIP panels are installed on the dividing wall between the two caves and on the rear wall of the detector cave. The positions of the polarizer vessel, reentrant vessel and beamstop are highlighted. (b) Section reference view of the detector cave showing ZHIP panels installed on the underside of the access well. (c) The cylindrical detector support frame consists of a detector cage mounted on stand legs. ZHIP panels will be used to block the direct line of sight between the detector and the support legs.}
\end{figure}

\begin{figure*}
\includegraphics[width=0.8\textwidth]{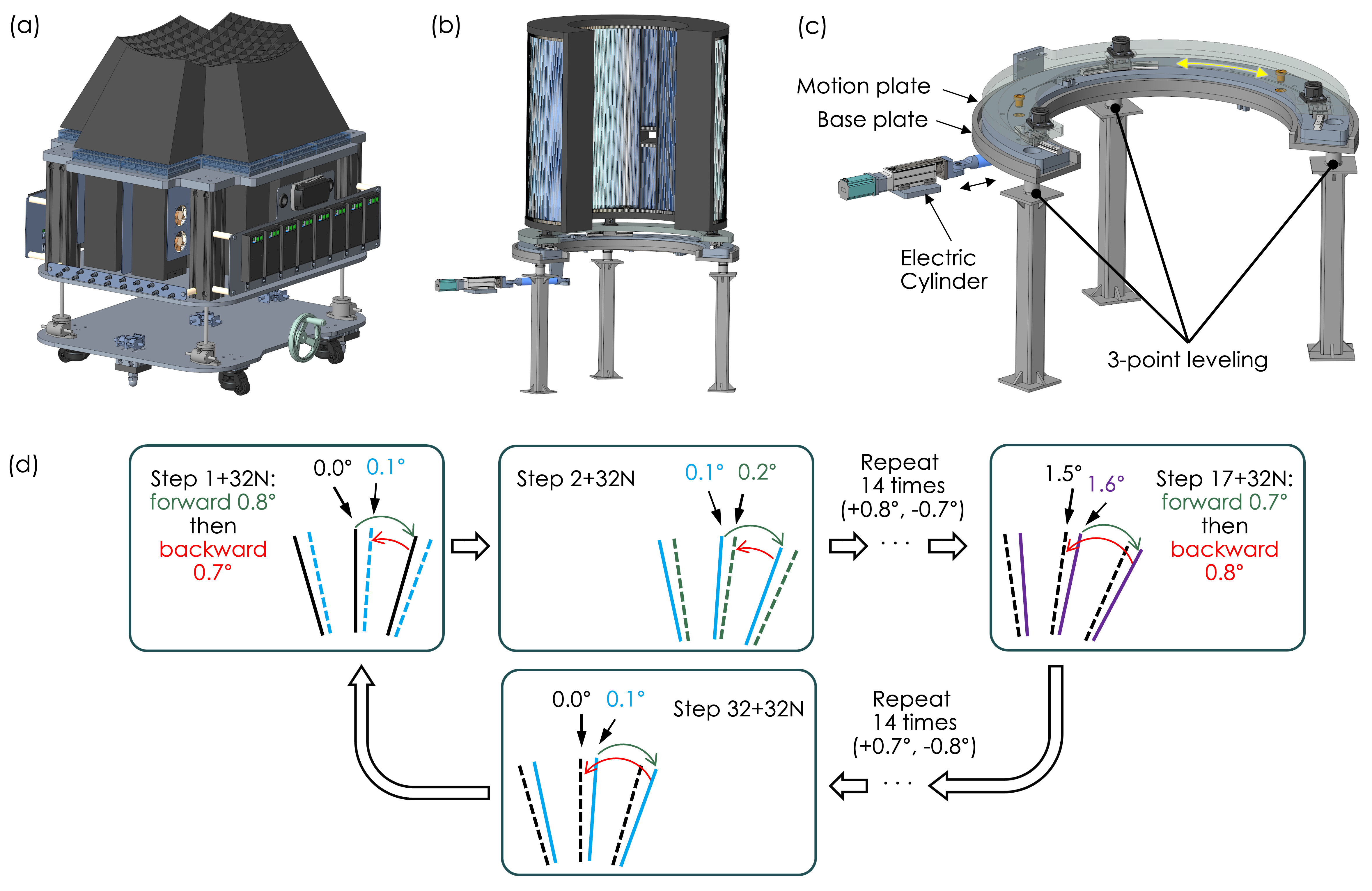}
\caption{\label{fig:cols}(a) A fixed multi-cone collimator for the bottom detector. (b) An oscillating radial collimator (ORC) for the vertical cylindrical detector. (c) The ORC's motion system enables axial alignment of curved guides, as indicated by the yellow line. The base plate is supported by a three-point leveling system for precise alignment. (d) Illustration of the ORC's shift mode operation: each cycle consists of 32 steps, with each step involving a forward and backward rotation. The solid lines represent the blade position before each step, and the dashed lines show the position after each step.}
\end{figure*}

The detector cave is located at the end of the instrument. Pioneer's detector will employ magnetic-field-insensitive silicon photomultiplier (SiPM) camera technology with a spatial resolution of 0.7~mm or better~\cite{loyd2024sub}, supporting experiments involving high-field magnets and enabling the instrument's high-resolution mode to probe large length scales beyond 100~\AA. A beamstop is located at the end of the detector cave to intercept and stop radiation remaining in the beam path after the sample to reduce the dose rate in the instrument hall to a safe level below 0.25~mrem/h when the source is running at the full capacity~\cite{STSConcept2020}.  Due to the spatial constraint, high density concrete is selected over the normal concrete for the beamstop.  Our preliminary neutronics calculations show that a beamstop of 1.85-m thick~$\times$~$0.80$-m wide~$\times$~$0.8$-m tall is sufficient. Because the nuclear interactions in the beamstop may result in backscattering of radiation towards the detector, an evacuated 0.6-m get-lost tube is installed through the back wall of the cave before the beamstop. This reduces the solid angle available to backscattering radiation, effectively reducing the background near the end station. A $0.05$-m B$_4$C plate will be installed at the end of the get-lost tube to absorb slow neutrons entering or backscattering out of the beamstop. 

Pioneer aims for measuring weak scattering signals, therefore effective shielding around the end station is critical to reduce background~\cite{gallmeier2006neutron, stone2019characterization}. The support frame of the cylindrical detector, as shown in Fig.~\ref{fig:B4C}(c), consists of an aluminum detector cage supported by a stainless steel stand. As most experiments at Pioneer will involve large sample environments, such as cryostats or cryomagnets, scattering beam collimators will be used to block the background scattering from reaching the detector~\cite{stone2015arcs}. Figure.~\ref{fig:cols}(a) shows the design of the 3D-printed conical collimator for the bottom flat detector~\cite{haberl20213d}, while the vertical cylindrical detector will use an ORC~\cite{stone2014radial}, as shown in Fig.~\ref{fig:cols}(b). Further details on the shielding panels and the ORC are provided below. 

\subsection{B$_4$C absorbing panels}{\label{sec:ZHIP}}
Pioneer will utilize B$_4$C absorbing panels near the end station to reduce background. As shown in Fig.\ref{fig:B4C}(a), the rear wall of the polarizer cave is lined with ZHIP panels to block neutrons leaked from the polarizer vessel and the guide, preventing them from reaching the detector. Similarly, the rear wall of the detector cave is covered with ZHIP panels to minimize backscattering from neutrons striking the wall. Additionally, ZHIP panels are used to shield the access well and the legs of the detector support structures, as shown in Figs.~\ref{fig:B4C}(b) and (c), respectively. ZHIP panels beneath the access well will further reduce ambient radiation levels above the cave, ensuring safe access for personnel.

\subsection{Detector collimators}{\label{sec:Col}}
\begin{figure}
\includegraphics[width=0.5\textwidth]{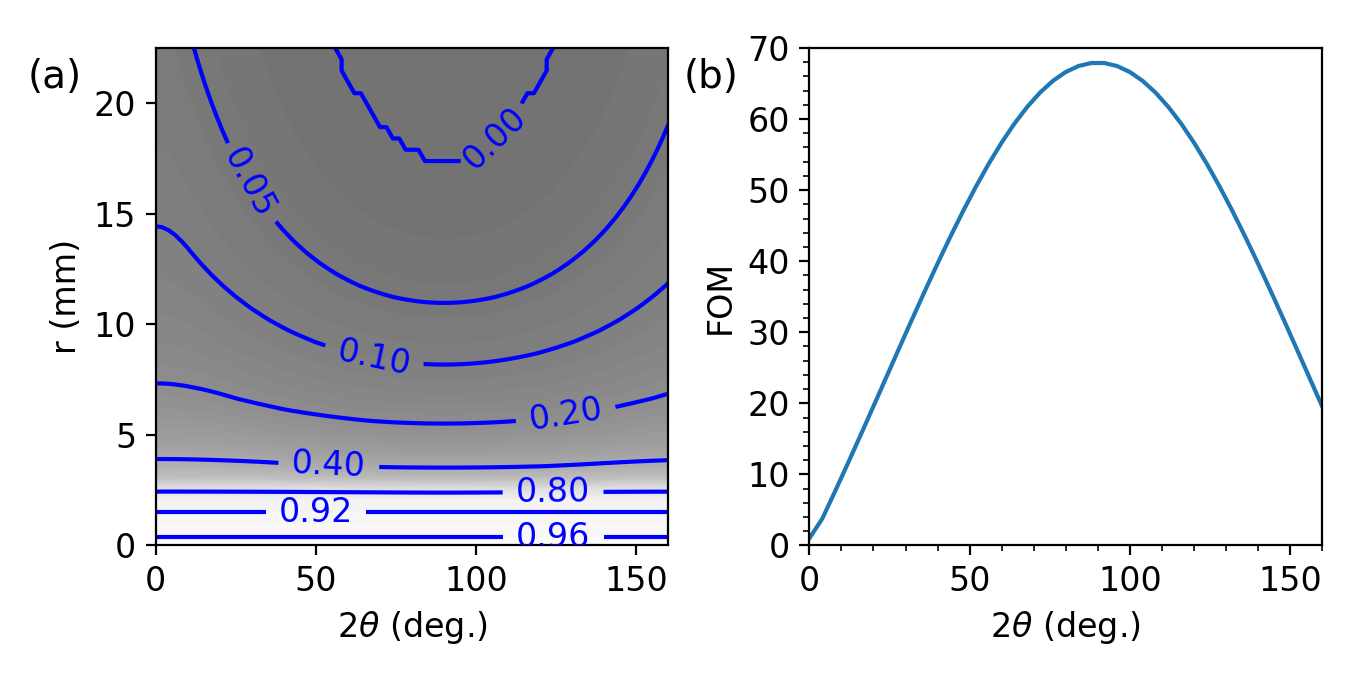}
\caption{\label{fig:col_FOM}(a) The in-plane visibility function for the ORC as a function of the in-plane detector angle ($2\theta$) and the distance ($r$) from the nominal sample center, assuming a beam size of 5~mm. (b) The figure of merit (FOM) for the ORC as a function of $2\theta$. See text for more details.}
\end{figure}

An oscillating radial collimator will be installed between the reentrant vessel and the cylindrical detector array. Pioneer's ORC will provide in-plane coverage from -100$^\circ$ to 160$^\circ$, and out-of-plane coverage from -36$^\circ$ to 47$^\circ$. The collimator's blades will have inner and outer radii of $R_1 = 475$ mm and $R_2 = 775$ mm, respectively, constrained by the detector and reentrant vessel. We have selected a blade separation of $2\theta = 0.8^\circ$, resulting in an impact length of $b = (R_1 R_2)/(R_2 - R_1) = 27.7$ mm~\cite{copley1994analysis}. The OSC design has considered the major sample environments anticipated for Pioneer, including closed-cycle refrigerators (60-90~mm diameter sample opening), liquid helium cryostats (60-90~mm diameter sample opening), and superconducting magnets (30-40~mm diameter sample opening)~\cite{SEs}. The 0.8$^\circ$ blade separation is chosen for simplicity rather than full optimization. The resulting impact parameter (27.7~mm) is comparable to values from previous instruments using similar sample environments~\cite{stone2014radial}. As shown in Fig.~\ref{fig:cols}(b), two blades are removed at low angles to install supporting plates, ensuring the structural reliability of this large collimator. A hole is placed in the direct beam path to avoid background scattering from the blades.

The figure-of-merit (FOM) of a radial collimator can be evaluated from the ratio of the quality factors with and without a collimator~\cite{copley1994analysis}. Assuming a neutron beam with a width of $2W$, the quality factor $Q$ is the ratio of the integrated visibility $V$ from the region within the beam ($0 < r < W$) and the area beyond ($W < r < R_1$), where $r$ is the distance from the sample center. The visibility represents the average view of the sample.

We have evaluated the performance of Pioneer's ORC, following the work by Copley and Cook and considering only the in-plane scattering angle~\cite{copley1994analysis}. Figure~\ref{fig:col_FOM}(a) shows the estimated visibility, $V_{W}(r, 2\theta)$, as a function of the in-plane detector angle $2\theta$ and the distance ($r$) from the nominal sample center~\cite{copley1994analysis}. The calculations assume a beam size of 5~mm and a constant blade thickness of 0.2~mm. Figure~\ref{fig:vis} shows visibility curves at selected detector angles with and without a collimator. As expected, the collimator reduces the sample's visibility and works best for detectors at 90$^\circ$. The quality factor $Q$ and FOM are defined as, 
\[ Q_W(2\theta) = \int_0^{W} r V_W(r, 2\theta) dr / \int_{W}^{R_1} r V_W(r, 2\theta) dr\] and 
\[ FOM(2\theta) = Q_W(2\theta)/Q_W^{nc} ,\] 
where $Q_W(2\theta)$ and $Q_W^{nc}$ are the quality factors with and without the radial collimator, respectively. Figure~\ref{fig:col_FOM}(b) shows that the FOM peaks at a value of 68 at 90$^{\circ}$, indicating a significant improvement in the signal-to-noise ratio due to the collimator.

Figures~\ref{fig:cols}(b) and (c) show the ORC and its motion system. The system includes a base plate and two independently adjustable motion plates, designed to ensure precise axial alignment of the curved guides. A new programmable drive concept combines an electric cylinder with a mechanical linkage to convert linear motion into stepped radial oscillation. The linkage allows remote motor positioning, minimizing interference from sample environments, particularly stray magnetic fields. Traditional ORC operations can cause periodic intensity fluctuations due to the dead time during blade direction changes, known as the shadow effect. This can be mitigated through a shift mode operation, where the blade movement angles is strategically adjusted, as illustrated in Fig.~\ref{fig:cols}(d). This approach ensures a more even distribution of shadows over time, and previous studies have shown the performance benefits of shift mode operations~\cite{nakamura2018performances}.

\section{Discussion}{\label{sec:discussion}}

To optimize beam transport performance, we have implemented strategies to reduce materials in the incident beam path. Pioneer will utilize a second disk chopper to solve the frame overlap issue. Although a frame overlap mirror offers a more cost-effective and lower-maintenance alternative, it causes flux attenuation and may increase phase space inhomogeneity. Beam monitors are only used for commissioning and diagnostics. Therefore, to prevent flux loss from beam monitors~\cite{issa2017characterization}, all monitors, except the one located after the sample, are retractable and remotely controlled. Additionally, efforts have been made to reduce the number of windows along the beam path, especially near the end station, where gate valves are used when necessary. Despite these efforts, the current design includes 23 windows, and we are investigating optimal materials to minimize flux loss while maintaining structural integrity.

A translatable system to switch between different guides is a common approach for controlling beam divergence. However, this option introduces strict alignment requirements due to Pioneer's small beam cross-section. Therefore, we have chosen the slit package option to prevent misalignment effects associated with translatable guides. Pioneer's guide system will also incorporate large gaps to accommodate essential optical components. When these components are moved out of the beam path, oversized in-guide, shielding collimators will be inserted, eliminating the need for precise alignment. These collimators will aid in blocking unwanted neutrons. However, additional neutronics studies are required to validate the selected materials and evaluate their efficacy in mitigating ambient background levels near the end station.

Considering Pioneer's operational wavelength range, we are considering the commercially available Gd$_2$O$_3$-coated blades for the ORC. These blades offer better attenuation performance compared to B$_4$C-based blades for neutrons with wavelengths $\geq 1$\AA~\cite{stone2014radial, nakamura2015oscillating, stone2019characterization}. Furthermore, a recent study suggests that GdZr alloy blades provide enhanced mechanical and neutronic properties over traditional Gd$_2$O$_3$-coated Mylar foils~\cite{qiu2022neutron}, presenting a promising alternative.

\section{Summary}{\label{sec:summary}}
We report on the progress in the optical design of Pioneer, focusing on maximizing neutron transport and control performance, reducing background, and simplifying engineering complexity. The rationale for the placement of major optical components, including the T$_{0}$ and disk choppers, shutter, and polarizing V-cavity, has been outlined. Given the small beam cross-section, we opted not to use translatable guides, thereby preventing issues from guide misalignment and enhancing operational reliability. Instead, when components such as the maintenance shield, T$_{0}$ chopper, shutter, and polarizing V-cavity are translated out of the beam path, oversized in-guide collimators will be inserted at the gaps to block unwanted neutrons. Pioneer will utilize slit packages to control beam size and divergence, with strategically positioned B$_4$C absorbing panels further reducing neutron background near the end station. Additionally, scattering beam collimators will be used to block undesired scattering from large sample environments. Pioneer's ORC will operate in a shift mode to minimize shadowing effects. While the optical design of Pioneer is the primary focus, the discussed principles and strategies are applicable to other neutron scattering instruments that use small beams and are designed for measuring weak signals in versatile sample environments.

\section*{Supplementary Material}
The supplementary materials include five figures. Figure~\ref{fig:TOFRes} shows the wavelength-dependent resolution at multiple moderator-to-sample distances (Sec.~\ref{sec:layout}). Figure~\ref{fig:2BLchoppers} illustrates the wavelength band using one disk chopper and two disk choppers, while Fig.~\ref{fig:BLchop_loc} presents possible locations of the two disk choppers. Figure~\ref{fig:DDchoice} compares the effective bandwidths of a single-disk versus a double-disk bandwidth-limiting chopper (Sec.~\ref{sec:chopper}). Lastly, Fig.~\ref{fig:vis} shows in-plane visibility curves at different detector angles (Sec.~\ref{sec:endstation}).

\begin{acknowledgments}
The authors thank Van Graves, David Anderson, Leighton Coates, and Kenneth Herwig for their valuable discussions. This research used resources of the Spallation Neutron Source Second Target Station Project at Oak Ridge National Laboratory (ORNL). ORNL is managed by UT-Battelle, LLC, for the U.S. Department of Energy's (DOE) Office of Science, the largest supporter of basic research in the physical sciences in the United States. This manuscript has been authored by UT-Battelle, LLC, under contract DE-AC05-00OR22725 with the U.S. Department of Energy (DOE), Basic Energy Sciences.
\end{acknowledgments}

\section*{Data Availability Statement}
The data that support the findings of this study are available from the corresponding author upon reasonable request.

\pagebreak
\renewcommand{\thefigure}{S\arabic{figure}}
\setcounter{figure}{0}

\begin{CJK*}{GB}{} 
\title[]{Supplementary Material: Optical design for the single crystal neutron diffractometer Pioneer}

\author{Yaohua Liu*}
\email[Author to whom correspondence should be addressed: ]{liuyh@ornl.gov}
\affiliation{Second Target Station, Oak Ridge National Laboratory, Oak Ridge, Tennessee 37831, USA}%

\author{Peter Torres}
\affiliation{Second Target Station, Oak Ridge National Laboratory, Oak Ridge, Tennessee 37831, USA}%

\author{Scott Dixon}
\affiliation{Second Target Station, Oak Ridge National Laboratory, Oak Ridge, Tennessee 37831, USA}%

\author{Cameron Hart} 
\affiliation{Second Target Station, Oak Ridge National Laboratory, Oak Ridge, Tennessee 37831, USA}%

\author{Darian Kent}
\affiliation{Second Target Station, Oak Ridge National Laboratory, Oak Ridge, Tennessee 37831, USA}%

\author{Anton Khaplanov} 
\affiliation{Second Target Station, Oak Ridge National Laboratory, Oak Ridge, Tennessee 37831, USA}%

\author{Bill M$^\textrm{c}$Hargue} 
\affiliation{Second Target Station, Oak Ridge National Laboratory, Oak Ridge, Tennessee 37831, USA}%

\author{Kumar Mohindroo} 
\affiliation{Second Target Station, Oak Ridge National Laboratory, Oak Ridge, Tennessee 37831, USA}%

\author{Rudolf Thermer} 
\affiliation{Second Target Station, Oak Ridge National Laboratory, Oak Ridge, Tennessee 37831, USA}%

\date{\today}
\maketitle
\end{CJK*}

\begin{figure*}
\includegraphics[width=0.5\textwidth]{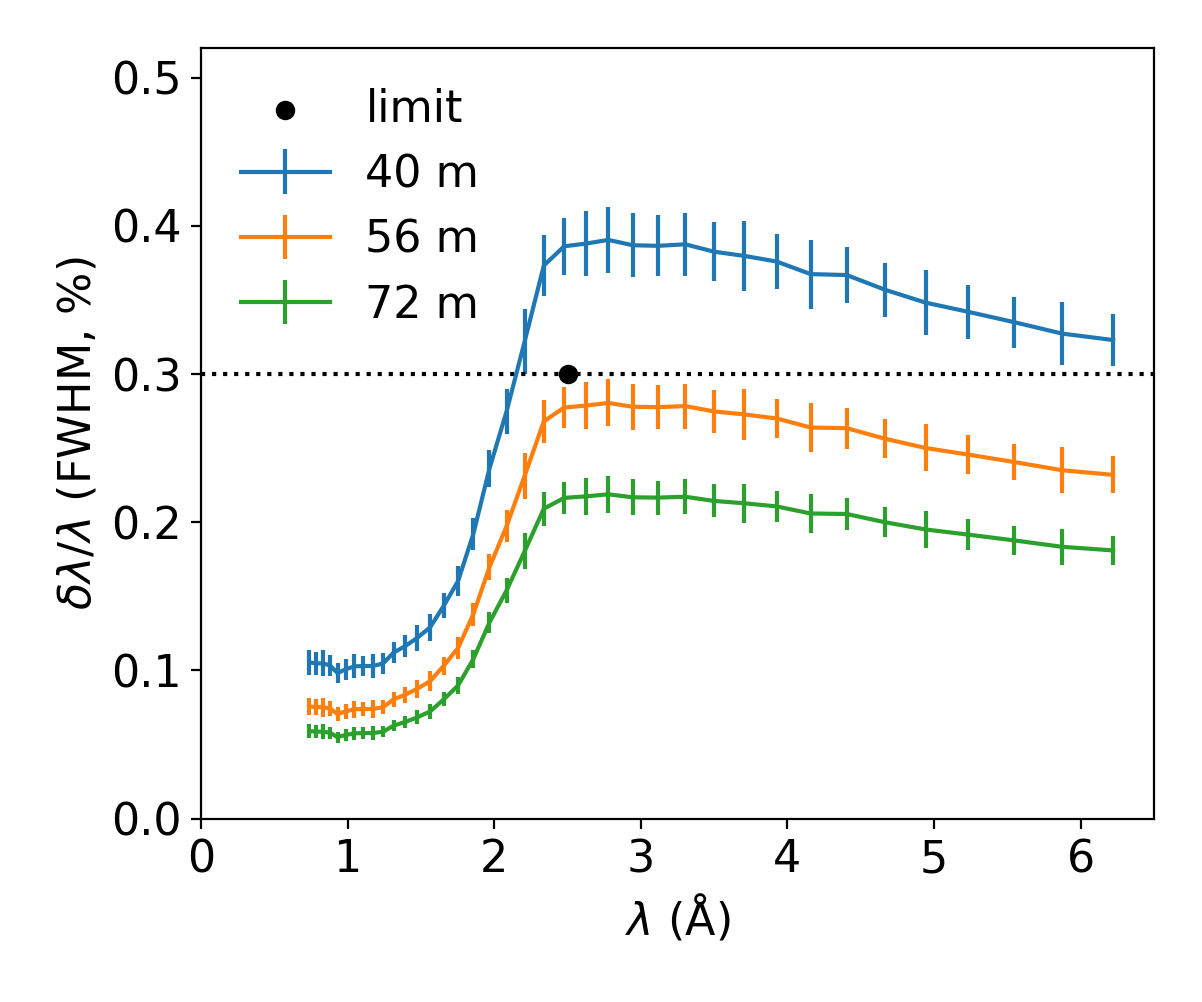}
\caption{\label{fig:TOFRes} Relative wavelength resolution at various moderator-to-sample distances. A constant sample-to-detector distance of 0.8-m is used for all cases. The calculation used the moderator source file of \emph{ST-13-CY46D-MID-March\_2022.dat}, provided by the STS neutronics group.}
\end{figure*}

\begin{figure*}
\includegraphics[width=0.5\textwidth]{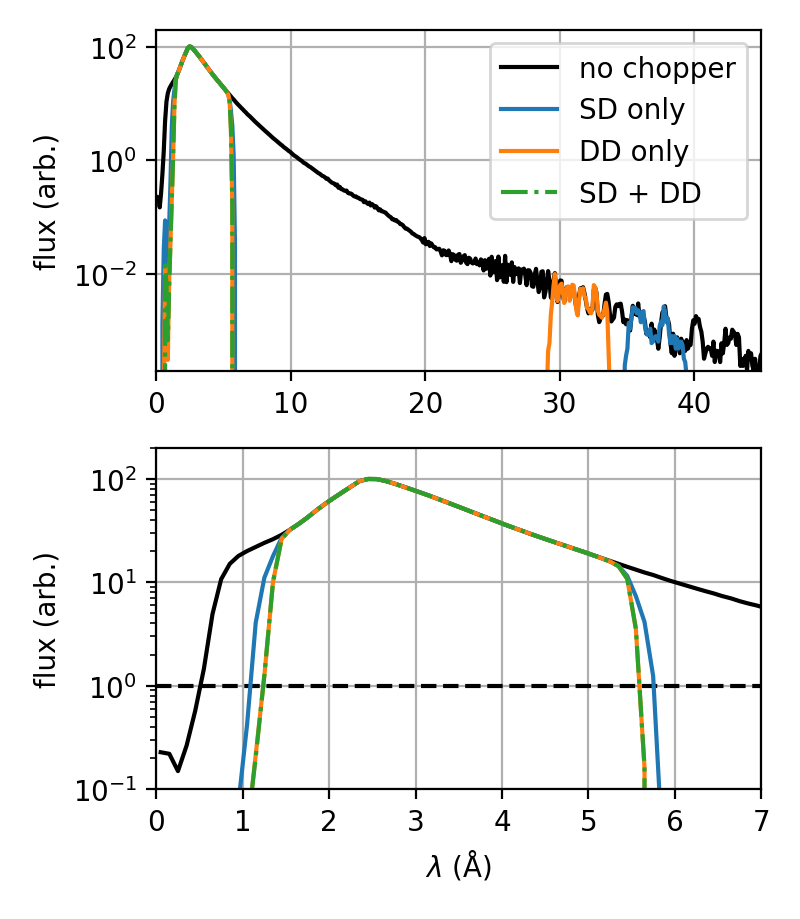}
\caption{\label{fig:2BLchoppers}Spectra from Monte Carlo ray tracing simulations. (a) Wavelength-dependent flux of the direct beam at the detector location ($L_{source-moderator}=56.8$~m) using various bandwidth-limiting chopper settings. Long-wavelength neutrons leak when only a single bandwidth-limiting chopper is used, either a single-disk chopper at 7.78~m or a double-disk chopper at 9.38~m. The leaked bands are centered around 31~\AA~and 37~\AA, respectively. The leakage can be avoided using both choppers. The fast fluctuations in the long-wavelength range are artifacts from coarse binning in the source file. (b) A zoomed-in view of the short-wavelength range. The dashed line indicates where the flux is at 1\% of the peak flux level. The spectra without bandwidth-limiting choppers are shown for comparison. The sharp flux drop below 0.8~\AA~is due to the rapidly reduced neutron transport efficiency, because the guide coating profile is optimized for $\ge 1.0$~\AA~neutrons. The simulations have assumed standardized STS large disk choppers (diameter = 1.24~m, opening height = 7~cm) with 100\% absorption efficiency, and have ignored the T$_0$ chopper.}
\end{figure*}

\begin{figure*}
\includegraphics[width=0.6\textwidth]{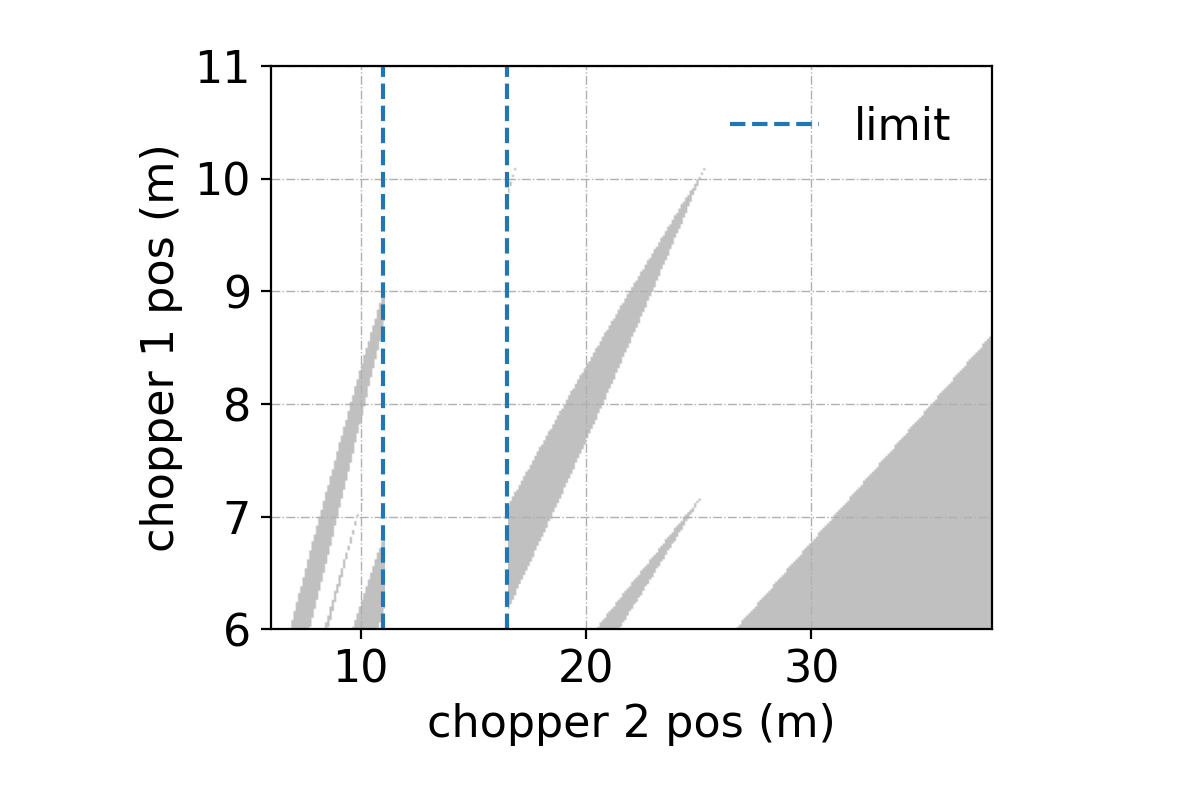}
\caption{\label{fig:BLchop_loc} The gray area indicates combinations of working locations of the two disk choppers for Pioneer, with a moderator-to-detector distance of 56.8~m and a source frequency of 15~Hz. The calculation has assumed a chopper frequency matching the source frequency (15~Hz) and ignored the neutron emission time. The region between the dashed lines is impractical due to constraints imposed by the STS bunker structure.}
\end{figure*}

\begin{figure*}
\includegraphics[width=0.75\textwidth]{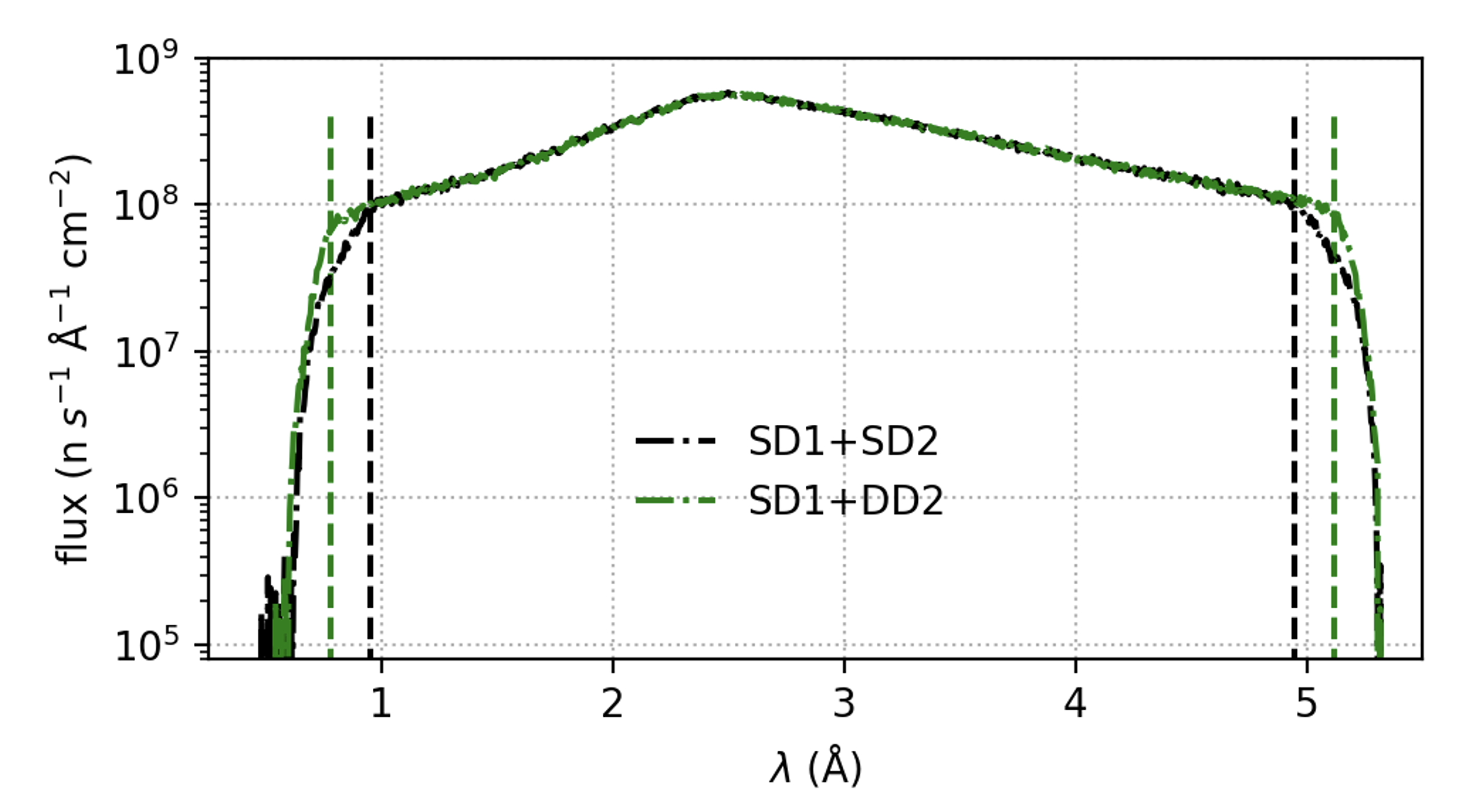}
\caption{\label{fig:DDchoice} Simulated spectra with various chopper settings show an effective bandwidth increase from 4.0 to 4.3~\AA~when the bandwidth-limiting chopper (second disk chopper) is switched from a single-disk to a double-disk large chopper, while the first remains a single-disk chopper. Chopper apertures are adjusted in each case to prevent frame overlap issues.}
\end{figure*}

\begin{figure*}
\includegraphics[width=0.5\textwidth]{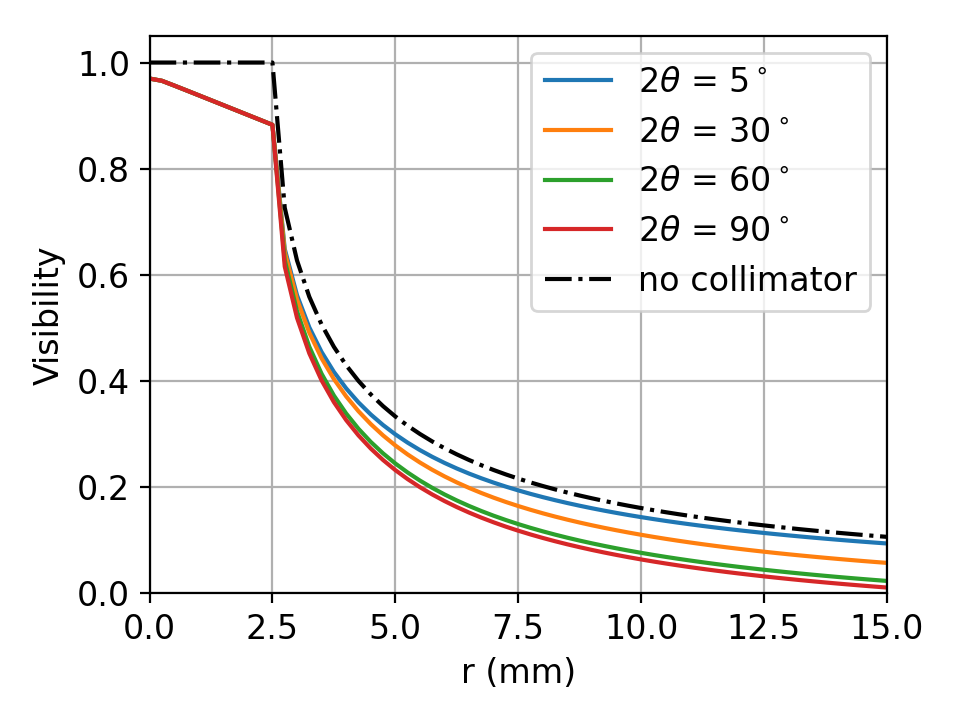}
\caption{\label{fig:vis} In-plane visibility curves of Pioneer's ORC at different detector angles. The visibility curve without a collimator is also shown for comparison, which is independent of the detector angle. The calculations have assumed a direct beam size of 5~mm. As expected, an ORC will reduce the sample's visibility. }
\end{figure*}


\begin{thebibliography}{36}%
\makeatletter
\providecommand \@ifxundefined [1]{%
 \@ifx{#1\undefined}
}%
\providecommand \@ifnum [1]{%
 \ifnum #1\expandafter \@firstoftwo
 \else \expandafter \@secondoftwo
 \fi
}%
\providecommand \@ifx [1]{%
 \ifx #1\expandafter \@firstoftwo
 \else \expandafter \@secondoftwo
 \fi
}%
\providecommand \natexlab [1]{#1}%
\providecommand \enquote  [1]{``#1''}%
\providecommand \bibnamefont  [1]{#1}%
\providecommand \bibfnamefont [1]{#1}%
\providecommand \citenamefont [1]{#1}%
\providecommand \href@noop [0]{\@secondoftwo}%
\providecommand \href [0]{\begingroup \@sanitize@url \@href}%
\providecommand \@href[1]{\@@startlink{#1}\@@href}%
\providecommand \@@href[1]{\endgroup#1\@@endlink}%
\providecommand \@sanitize@url [0]{\catcode `\\12\catcode `\$12\catcode
  `\&12\catcode `\#12\catcode `\^12\catcode `\_12\catcode `\%12\relax}%
\providecommand \@@startlink[1]{}%
\providecommand \@@endlink[0]{}%
\providecommand \url  [0]{\begingroup\@sanitize@url \@url }%
\providecommand \@url [1]{\endgroup\@href {#1}{\urlprefix }}%
\providecommand \urlprefix  [0]{URL }%
\providecommand \Eprint [0]{\href }%
\providecommand \doibase [0]{http://dx.doi.org/}%
\providecommand \selectlanguage [0]{\@gobble}%
\providecommand \bibinfo  [0]{\@secondoftwo}%
\providecommand \bibfield  [0]{\@secondoftwo}%
\providecommand \translation [1]{[#1]}%
\providecommand \BibitemOpen [0]{}%
\providecommand \bibitemStop [0]{}%
\providecommand \bibitemNoStop [0]{.\EOS\space}%
\providecommand \EOS [0]{\spacefactor3000\relax}%
\providecommand \BibitemShut  [1]{\csname bibitem#1\endcsname}%
\let\auto@bib@innerbib\@empty
\bibitem [{\citenamefont {Carpenter}\ and\ \citenamefont
  {Loong}(2015)}]{carpenter2015elements}%
  \BibitemOpen
  \bibfield  {author} {\bibinfo {author} {\bibfnamefont {J.~M.}\ \bibnamefont
  {Carpenter}}\ and\ \bibinfo {author} {\bibfnamefont {C.-K.}\ \bibnamefont
  {Loong}},\ }\href@noop {} {\emph {\bibinfo {title} {Elements of slow-neutron
  scattering}}}\ (\bibinfo  {publisher} {Cambridge University Press},\ \bibinfo
  {year} {2015})\BibitemShut {NoStop}%
\bibitem [{\citenamefont {ORNL}(2020)}]{adams2020first}%
  \BibitemOpen
  \bibfield  {author} {\bibinfo {author} {\bibnamefont {ORNL}},\ }\href
  {https://neutrons.ornl.gov/sites/default/files/STS_First_Experiments_Report.pdf}
  {\emph {\bibinfo {title} {{First Experiments: New Science Opportunities at
  the Spallation Neutron Source Second Target Station (abridged)}}}},\ \bibinfo
  {type} {Tech. Rep.}\ (\bibinfo  {institution} {Oak Ridge National Lab., Oak
  Ridge, TN (United States)},\ \bibinfo {year} {2020})\BibitemShut {NoStop}%
\bibitem [{STS(2020)}]{STSConcept2020}%
  \BibitemOpen
  \href {https://neutrons.ornl.gov/sites/default/files/STS_CDR_Vol1_v2.pdf}
  {\emph {\bibinfo {title} {{Spallation Neutron Source Second Target Station
  Conceptual Design Report Volume 1: Overview, Technical and Experiment
  Systems}}}},\ \bibinfo {type} {Technical Report}\ \bibinfo {number}
  {S01010000-TR0001, R00}\ (\bibinfo  {institution} {Oak Ridge National
  Laboratory},\ \bibinfo {year} {March 2020})\BibitemShut {NoStop}%
\bibitem [{\citenamefont {Liu}\ \emph {et~al.}(2022)\citenamefont {Liu},
  \citenamefont {Cao}, \citenamefont {Rosenkranz}, \citenamefont {Frost},
  \citenamefont {Huegle}, \citenamefont {Lin}, \citenamefont {Torres},
  \citenamefont {Stoica},\ and\ \citenamefont {Chakoumakos}}]{liu2022Pioneer}%
  \BibitemOpen
  \bibfield  {author} {\bibinfo {author} {\bibfnamefont {Y.}~\bibnamefont
  {Liu}}, \bibinfo {author} {\bibfnamefont {H.}~\bibnamefont {Cao}}, \bibinfo
  {author} {\bibfnamefont {S.}~\bibnamefont {Rosenkranz}}, \bibinfo {author}
  {\bibfnamefont {M.}~\bibnamefont {Frost}}, \bibinfo {author} {\bibfnamefont
  {T.}~\bibnamefont {Huegle}}, \bibinfo {author} {\bibfnamefont {J.~Y.}\
  \bibnamefont {Lin}}, \bibinfo {author} {\bibfnamefont {P.}~\bibnamefont
  {Torres}}, \bibinfo {author} {\bibfnamefont {A.}~\bibnamefont {Stoica}}, \
  and\ \bibinfo {author} {\bibfnamefont {B.~C.}\ \bibnamefont {Chakoumakos}},\
  }\href@noop {} {\bibfield  {journal} {\bibinfo  {journal} {Review of
  Scientific Instruments}\ }\textbf {\bibinfo {volume} {93}},\ \bibinfo {pages}
  {073901} (\bibinfo {year} {2022})}\BibitemShut {NoStop}%
\bibitem [{\citenamefont {Liu}(2024)}]{liu2024general}%
  \BibitemOpen
  \bibfield  {author} {\bibinfo {author} {\bibfnamefont {Y.}~\bibnamefont
  {Liu}},\ }\href {\doibase 10.1063/5.0212920} {\bibfield  {journal} {\bibinfo
  {journal} {Review of Scientific Instruments}\ }\textbf {\bibinfo {volume}
  {95}},\ \bibinfo {pages} {073902} (\bibinfo {year} {2024})}\BibitemShut
  {NoStop}%
\bibitem [{\citenamefont {Liu}\ and\ \citenamefont
  {Torres}(2025)}]{liu2025incident}%
  \BibitemOpen
  \bibfield  {author} {\bibinfo {author} {\bibfnamefont {Y.}~\bibnamefont
  {Liu}}\ and\ \bibinfo {author} {\bibfnamefont {P.}~\bibnamefont {Torres}},\
  }\href {\doibase 10.1063/5.0259079} {\bibfield  {journal} {\bibinfo
  {journal} {Review of Scientific Instruments}\ }\textbf {\bibinfo {volume}
  {96}},\ \bibinfo {pages} {in press} (\bibinfo {year} {2025})}\BibitemShut
  {NoStop}%
\bibitem [{\citenamefont {Willendrup}\ \emph {et~al.}(2014)\citenamefont
  {Willendrup}, \citenamefont {Farhi}, \citenamefont {Knudsen}, \citenamefont
  {Filges},\ and\ \citenamefont {Lefmann}}]{willendrup2014mcstas}%
  \BibitemOpen
  \bibfield  {author} {\bibinfo {author} {\bibfnamefont {P.}~\bibnamefont
  {Willendrup}}, \bibinfo {author} {\bibfnamefont {E.}~\bibnamefont {Farhi}},
  \bibinfo {author} {\bibfnamefont {E.}~\bibnamefont {Knudsen}}, \bibinfo
  {author} {\bibfnamefont {U.}~\bibnamefont {Filges}}, \ and\ \bibinfo {author}
  {\bibfnamefont {K.}~\bibnamefont {Lefmann}},\ }\href@noop {} {\bibfield
  {journal} {\bibinfo  {journal} {Journal of Neutron Research}\ }\textbf
  {\bibinfo {volume} {17}},\ \bibinfo {pages} {35} (\bibinfo {year}
  {2014})}\BibitemShut {NoStop}%
\bibitem [{\citenamefont {Lin}\ \emph {et~al.}(2016)\citenamefont {Lin},
  \citenamefont {Smith}, \citenamefont {Granroth}, \citenamefont {Abernathy},
  \citenamefont {Lumsden}, \citenamefont {Winn}, \citenamefont {Aczel},
  \citenamefont {Aivazis},\ and\ \citenamefont {Fultz}}]{lin2016mcvine}%
  \BibitemOpen
  \bibfield  {author} {\bibinfo {author} {\bibfnamefont {J.~Y.}\ \bibnamefont
  {Lin}}, \bibinfo {author} {\bibfnamefont {H.~L.}\ \bibnamefont {Smith}},
  \bibinfo {author} {\bibfnamefont {G.~E.}\ \bibnamefont {Granroth}}, \bibinfo
  {author} {\bibfnamefont {D.~L.}\ \bibnamefont {Abernathy}}, \bibinfo {author}
  {\bibfnamefont {M.~D.}\ \bibnamefont {Lumsden}}, \bibinfo {author}
  {\bibfnamefont {B.}~\bibnamefont {Winn}}, \bibinfo {author} {\bibfnamefont
  {A.~A.}\ \bibnamefont {Aczel}}, \bibinfo {author} {\bibfnamefont
  {M.}~\bibnamefont {Aivazis}}, \ and\ \bibinfo {author} {\bibfnamefont
  {B.}~\bibnamefont {Fultz}},\ }\href@noop {} {\bibfield  {journal} {\bibinfo
  {journal} {Nuclear Instruments and Methods in Physics Research Section A:
  Accelerators, Spectrometers, Detectors and Associated Equipment}\ }\textbf
  {\bibinfo {volume} {810}},\ \bibinfo {pages} {86} (\bibinfo {year}
  {2016})}\BibitemShut {NoStop}%
\bibitem [{STS(2024)}]{STSICS2024}%
  \BibitemOpen
  \href@noop {} {\emph {\bibinfo {title} {Second Target Station Project:
  Integrated Control Systems and Instrument Systems Chopper Naming
  Procedure}}},\ \bibinfo {type} {Technical Report}\ \bibinfo {number}
  {S06040200-PCD10000-R00}\ (\bibinfo  {institution} {Oak Ridge National
  Laboratory},\ \bibinfo {year} {2024})\BibitemShut {NoStop}%
\bibitem [{\citenamefont {Mezei}(1989)}]{mezei1989very}%
  \BibitemOpen
  \bibfield  {author} {\bibinfo {author} {\bibfnamefont {F.}~\bibnamefont
  {Mezei}},\ }in\ \href@noop {} {\emph {\bibinfo {booktitle} {Thin Film Neutron
  Optical Devices: Mirrors, Supermirrors, Multilayer Monochromators,
  Polarizers, and Beam Guides}}},\ Vol.\ \bibinfo {volume} {983}\ (\bibinfo
  {organization} {SPIE},\ \bibinfo {year} {1989})\ pp.\ \bibinfo {pages}
  {10--17}\BibitemShut {NoStop}%
\bibitem [{\citenamefont {Fitzsimmons}\ and\ \citenamefont
  {Majkrzak}(2005)}]{fitzsimmons2005application}%
  \BibitemOpen
  \bibfield  {author} {\bibinfo {author} {\bibfnamefont {M.}~\bibnamefont
  {Fitzsimmons}}\ and\ \bibinfo {author} {\bibfnamefont {C.}~\bibnamefont
  {Majkrzak}},\ }\enquote {\bibinfo {title} {Application of polarized neutron
  reflectometry to studies of artificially structured magnetic materials},}\
  in\ \href@noop {} {\emph {\bibinfo {booktitle} {Modern Techniques for
  Characterizing Magnetic Materials}}},\ \bibinfo {editor} {edited by\ \bibinfo
  {editor} {\bibfnamefont {Y.}~\bibnamefont {Zhu}}}\ (\bibinfo  {publisher}
  {Springer},\ \bibinfo {year} {2005})\ pp.\ \bibinfo {pages}
  {107--155}\BibitemShut {NoStop}%
\bibitem [{ZHI()}]{ZHIP}%
  \BibitemOpen
  \href@noop {} {}\bibinfo {howpublished}
  {\url{https://www.dielectricsciences.com/neutron-absorbing-products}},\
  \bibinfo {note} {accessed: 2024-10-23}\BibitemShut {NoStop}%
\bibitem [{\citenamefont {Abernathy}\ \emph {et~al.}(2012)\citenamefont
  {Abernathy}, \citenamefont {Stone}, \citenamefont {Loguillo}, \citenamefont
  {Lucas}, \citenamefont {Delaire}, \citenamefont {Tang}, \citenamefont {Lin},\
  and\ \citenamefont {Fultz}}]{abernathy2012design}%
  \BibitemOpen
  \bibfield  {author} {\bibinfo {author} {\bibfnamefont {D.~L.}\ \bibnamefont
  {Abernathy}}, \bibinfo {author} {\bibfnamefont {M.~B.}\ \bibnamefont
  {Stone}}, \bibinfo {author} {\bibfnamefont {M.}~\bibnamefont {Loguillo}},
  \bibinfo {author} {\bibfnamefont {M.}~\bibnamefont {Lucas}}, \bibinfo
  {author} {\bibfnamefont {O.}~\bibnamefont {Delaire}}, \bibinfo {author}
  {\bibfnamefont {X.}~\bibnamefont {Tang}}, \bibinfo {author} {\bibfnamefont
  {J.}~\bibnamefont {Lin}}, \ and\ \bibinfo {author} {\bibfnamefont
  {B.}~\bibnamefont {Fultz}},\ }\href@noop {} {\bibfield  {journal} {\bibinfo
  {journal} {Review of Scientific Instruments}\ }\textbf {\bibinfo {volume}
  {83}},\ \bibinfo {pages} {15114} (\bibinfo {year} {2012})}\BibitemShut
  {NoStop}%
\bibitem [{\citenamefont {Jones}\ \emph {et~al.}(1987)\citenamefont {Jones},
  \citenamefont {Davidson}, \citenamefont {Boland}, \citenamefont {Bowden},\
  and\ \citenamefont {Taylor}}]{jones1987HET}%
  \BibitemOpen
  \bibfield  {author} {\bibinfo {author} {\bibfnamefont {T.}~\bibnamefont
  {Jones}}, \bibinfo {author} {\bibfnamefont {I.}~\bibnamefont {Davidson}},
  \bibinfo {author} {\bibfnamefont {B.}~\bibnamefont {Boland}}, \bibinfo
  {author} {\bibfnamefont {Z.}~\bibnamefont {Bowden}}, \ and\ \bibinfo {author}
  {\bibfnamefont {A.}~\bibnamefont {Taylor}},\ }in\ \href@noop {} {\emph
  {\bibinfo {booktitle} {Proceedings of the 9th Meeting of the International
  Collaboration on Advanced Neutron Sources, Swiss Institute for Nuclear
  Research}}}\ (\bibinfo {year} {1987})\ pp.\ \bibinfo {pages}
  {529--533}\BibitemShut {NoStop}%
\bibitem [{\citenamefont {Santoro}\ \emph {et~al.}(2022)\citenamefont
  {Santoro}, \citenamefont {Andersen}, \citenamefont {Khaplanov}, \citenamefont
  {Kolevatov}, \citenamefont {Gonzalez}, \citenamefont {Gruenauer},
  \citenamefont {Mag{\'a}n},\ and\ \citenamefont
  {Randriamalala}}]{Santoro2022}%
  \BibitemOpen
  \bibfield  {author} {\bibinfo {author} {\bibfnamefont {V.}~\bibnamefont
  {Santoro}}, \bibinfo {author} {\bibfnamefont {K.}~\bibnamefont {Andersen}},
  \bibinfo {author} {\bibfnamefont {A.}~\bibnamefont {Khaplanov}}, \bibinfo
  {author} {\bibfnamefont {R.}~\bibnamefont {Kolevatov}}, \bibinfo {author}
  {\bibfnamefont {O.}~\bibnamefont {Gonzalez}}, \bibinfo {author}
  {\bibfnamefont {F.}~\bibnamefont {Gruenauer}}, \bibinfo {author}
  {\bibfnamefont {M.}~\bibnamefont {Mag{\'a}n}}, \ and\ \bibinfo {author}
  {\bibfnamefont {T.}~\bibnamefont {Randriamalala}},\ }in\ \href {\doibase
  10.13182/ICRSRPSD22-39415} {\emph {\bibinfo {booktitle} {Proceedings of the
  14th International Conference on Radiation Shielding and 21st Topical Meeting
  of the Radiation Protection and Shielding Division, ICRS 2022/RPSD 2022}}}\
  (\bibinfo  {publisher} {American Nuclear Society},\ \bibinfo {year} {2022})\
  pp.\ \bibinfo {pages} {427--430}\BibitemShut {NoStop}%
\bibitem [{STS(2018)}]{STSChopper2018}%
  \BibitemOpen
  \href@noop {} {\emph {\bibinfo {title} {STS Instruments Conceptual Design 15
  Hz Chopper Models}}},\ \bibinfo {type} {Technical Report}\ \bibinfo {number}
  {STS04-41-IN0002-RA}\ (\bibinfo  {institution} {Oak Ridge National
  Laboratory},\ \bibinfo {year} {2018})\BibitemShut {NoStop}%
\bibitem [{\citenamefont {Itoh}\ \emph {et~al.}(2012)\citenamefont {Itoh},
  \citenamefont {Ueno}, \citenamefont {Ohkubo}, \citenamefont {Sagehashi},
  \citenamefont {Funahashi},\ and\ \citenamefont {Yokoo}}]{itoh2012t0}%
  \BibitemOpen
  \bibfield  {author} {\bibinfo {author} {\bibfnamefont {S.}~\bibnamefont
  {Itoh}}, \bibinfo {author} {\bibfnamefont {K.}~\bibnamefont {Ueno}}, \bibinfo
  {author} {\bibfnamefont {R.}~\bibnamefont {Ohkubo}}, \bibinfo {author}
  {\bibfnamefont {H.}~\bibnamefont {Sagehashi}}, \bibinfo {author}
  {\bibfnamefont {Y.}~\bibnamefont {Funahashi}}, \ and\ \bibinfo {author}
  {\bibfnamefont {T.}~\bibnamefont {Yokoo}},\ }\href@noop {} {\bibfield
  {journal} {\bibinfo  {journal} {Nuclear Instruments and Methods in Physics
  Research Section A: Accelerators, Spectrometers, Detectors and Associated
  Equipment}\ }\textbf {\bibinfo {volume} {661}},\ \bibinfo {pages} {86}
  (\bibinfo {year} {2012})}\BibitemShut {NoStop}%
\bibitem [{\citenamefont {Violini}\ \emph {et~al.}(2014)\citenamefont
  {Violini}, \citenamefont {Voigt},\ and\ \citenamefont
  {Br{\"u}ckel}}]{violini2014investigation}%
  \BibitemOpen
  \bibfield  {author} {\bibinfo {author} {\bibfnamefont {N.}~\bibnamefont
  {Violini}}, \bibinfo {author} {\bibfnamefont {J.}~\bibnamefont {Voigt}}, \
  and\ \bibinfo {author} {\bibfnamefont {T.}~\bibnamefont {Br{\"u}ckel}},\ }in\
  \href@noop {} {\emph {\bibinfo {booktitle} {Proceedings of the 21st Meeting
  of the International Collaboration on Advanced Neutron Sources (ICANS
  XXI)}}}\ (\bibinfo {year} {2014})\ pp.\ \bibinfo {pages}
  {272--277}\BibitemShut {NoStop}%
\bibitem [{\citenamefont {Wang}\ \emph {et~al.}(2023)\citenamefont {Wang},
  \citenamefont {Cai}, \citenamefont {Zhang}, \citenamefont {Guo},
  \citenamefont {Deng}, \citenamefont {Wang}, \citenamefont {Yin},
  \citenamefont {Xu}, \citenamefont {Wang}, \citenamefont {Liang} \emph
  {et~al.}}]{wang2023physical}%
  \BibitemOpen
  \bibfield  {author} {\bibinfo {author} {\bibfnamefont {P.}~\bibnamefont
  {Wang}}, \bibinfo {author} {\bibfnamefont {W.}~\bibnamefont {Cai}}, \bibinfo
  {author} {\bibfnamefont {Q.}~\bibnamefont {Zhang}}, \bibinfo {author}
  {\bibfnamefont {J.}~\bibnamefont {Guo}}, \bibinfo {author} {\bibfnamefont
  {L.}~\bibnamefont {Deng}}, \bibinfo {author} {\bibfnamefont {F.}~\bibnamefont
  {Wang}}, \bibinfo {author} {\bibfnamefont {W.}~\bibnamefont {Yin}}, \bibinfo
  {author} {\bibfnamefont {J.}~\bibnamefont {Xu}}, \bibinfo {author}
  {\bibfnamefont {S.}~\bibnamefont {Wang}}, \bibinfo {author} {\bibfnamefont
  {T.}~\bibnamefont {Liang}},  \emph {et~al.},\ }\href@noop {} {\bibfield
  {journal} {\bibinfo  {journal} {Nuclear Instruments and Methods in Physics
  Research Section A: Accelerators, Spectrometers, Detectors and Associated
  Equipment}\ }\textbf {\bibinfo {volume} {1055}},\ \bibinfo {pages} {168520}
  (\bibinfo {year} {2023})}\BibitemShut {NoStop}%
\bibitem [{\citenamefont {Carpenter}\ \emph {et~al.}(1984)\citenamefont
  {Carpenter}, \citenamefont {Lander},\ and\ \citenamefont
  {Windsor}}]{carpenter1984instrumentation}%
  \BibitemOpen
  \bibfield  {author} {\bibinfo {author} {\bibfnamefont {J.}~\bibnamefont
  {Carpenter}}, \bibinfo {author} {\bibfnamefont {G.}~\bibnamefont {Lander}}, \
  and\ \bibinfo {author} {\bibfnamefont {C.}~\bibnamefont {Windsor}},\
  }\href@noop {} {\bibfield  {journal} {\bibinfo  {journal} {Review of
  scientific instruments}\ }\textbf {\bibinfo {volume} {55}},\ \bibinfo {pages}
  {1019} (\bibinfo {year} {1984})}\BibitemShut {NoStop}%
\bibitem [{\citenamefont {Jauch}(1997)}]{jauch1997prospects}%
  \BibitemOpen
  \bibfield  {author} {\bibinfo {author} {\bibfnamefont {W.}~\bibnamefont
  {Jauch}},\ }\href@noop {} {\bibfield  {journal} {\bibinfo  {journal} {Journal
  of Neutron Research}\ }\textbf {\bibinfo {volume} {6}},\ \bibinfo {pages}
  {161} (\bibinfo {year} {1997})}\BibitemShut {NoStop}%
\bibitem [{\citenamefont {Zendler}\ \emph {et~al.}(2015)\citenamefont
  {Zendler}, \citenamefont {Rodriguez},\ and\ \citenamefont
  {Bentley}}]{zendler2015generic}%
  \BibitemOpen
  \bibfield  {author} {\bibinfo {author} {\bibfnamefont {C.}~\bibnamefont
  {Zendler}}, \bibinfo {author} {\bibfnamefont {D.~M.}\ \bibnamefont
  {Rodriguez}}, \ and\ \bibinfo {author} {\bibfnamefont {P.}~\bibnamefont
  {Bentley}},\ }\href@noop {} {\bibfield  {journal} {\bibinfo  {journal}
  {Nuclear Instruments and Methods in Physics Research Section A: Accelerators,
  Spectrometers, Detectors and Associated Equipment}\ }\textbf {\bibinfo
  {volume} {803}},\ \bibinfo {pages} {89} (\bibinfo {year} {2015})}\BibitemShut
  {NoStop}%
\bibitem [{\citenamefont {Lin}\ \emph {et~al.}(2023)\citenamefont {Lin},
  \citenamefont {Huegle}, \citenamefont {Coates},\ and\ \citenamefont
  {Stoica}}]{lin2023realistic}%
  \BibitemOpen
  \bibfield  {author} {\bibinfo {author} {\bibfnamefont {J.~Y.}\ \bibnamefont
  {Lin}}, \bibinfo {author} {\bibfnamefont {T.}~\bibnamefont {Huegle}},
  \bibinfo {author} {\bibfnamefont {L.}~\bibnamefont {Coates}}, \ and\ \bibinfo
  {author} {\bibfnamefont {A.~D.}\ \bibnamefont {Stoica}},\ }\href {\doibase
  https://doi.org/10.1016/j.nima.2022.167881} {\bibfield  {journal} {\bibinfo
  {journal} {Nuclear Instruments and Methods in Physics Research Section A:
  Accelerators, Spectrometers, Detectors and Associated Equipment}\ }\textbf
  {\bibinfo {volume} {1047}},\ \bibinfo {pages} {167881} (\bibinfo {year}
  {2023})}\BibitemShut {NoStop}%
\bibitem [{\citenamefont {Santoro}\ \emph {et~al.}(2018)\citenamefont
  {Santoro}, \citenamefont {DiJulio}, \citenamefont {Ansell}, \citenamefont
  {Cherkashyna}, \citenamefont {Muhrer},\ and\ \citenamefont
  {Bentley}}]{santoro2018study}%
  \BibitemOpen
  \bibfield  {author} {\bibinfo {author} {\bibfnamefont {V.}~\bibnamefont
  {Santoro}}, \bibinfo {author} {\bibfnamefont {D.}~\bibnamefont {DiJulio}},
  \bibinfo {author} {\bibfnamefont {S.}~\bibnamefont {Ansell}}, \bibinfo
  {author} {\bibfnamefont {N.}~\bibnamefont {Cherkashyna}}, \bibinfo {author}
  {\bibfnamefont {G.}~\bibnamefont {Muhrer}}, \ and\ \bibinfo {author}
  {\bibfnamefont {P.~M.}\ \bibnamefont {Bentley}},\ }in\ \href@noop {} {\emph
  {\bibinfo {booktitle} {Journal of Physics: Conference Series}}},\ Vol.\
  \bibinfo {volume} {1046}\ (\bibinfo {organization} {IOP Publishing},\
  \bibinfo {year} {2018})\ p.\ \bibinfo {pages} {012010}\BibitemShut {NoStop}%
\bibitem [{\citenamefont {Loyd}\ \emph {et~al.}(2024)\citenamefont {Loyd},
  \citenamefont {Khaplanov}, \citenamefont {Sedov}, \citenamefont {Beal},
  \citenamefont {Visscher}, \citenamefont {Donahue}, \citenamefont {Montcalm},
  \citenamefont {Warren}, \citenamefont {Butz}, \citenamefont {Boone} \emph
  {et~al.}}]{loyd2024sub}%
  \BibitemOpen
  \bibfield  {author} {\bibinfo {author} {\bibfnamefont {M.}~\bibnamefont
  {Loyd}}, \bibinfo {author} {\bibfnamefont {A.}~\bibnamefont {Khaplanov}},
  \bibinfo {author} {\bibfnamefont {V.}~\bibnamefont {Sedov}}, \bibinfo
  {author} {\bibfnamefont {J.}~\bibnamefont {Beal}}, \bibinfo {author}
  {\bibfnamefont {T.}~\bibnamefont {Visscher}}, \bibinfo {author}
  {\bibfnamefont {C.}~\bibnamefont {Donahue}}, \bibinfo {author} {\bibfnamefont
  {C.}~\bibnamefont {Montcalm}}, \bibinfo {author} {\bibfnamefont
  {G.}~\bibnamefont {Warren}}, \bibinfo {author} {\bibfnamefont
  {R.}~\bibnamefont {Butz}}, \bibinfo {author} {\bibfnamefont {C.}~\bibnamefont
  {Boone}},  \emph {et~al.},\ }\href@noop {} {\bibfield  {journal} {\bibinfo
  {journal} {Nuclear Instruments and Methods in Physics Research Section A:
  Accelerators, Spectrometers, Detectors and Associated Equipment}\ }\textbf
  {\bibinfo {volume} {1058}},\ \bibinfo {pages} {168871} (\bibinfo {year}
  {2024})}\BibitemShut {NoStop}%
\bibitem [{\citenamefont {Gallmeier}\ \emph {et~al.}(2006)\citenamefont
  {Gallmeier}, \citenamefont {Ferguson}, \citenamefont {Iverson}, \citenamefont
  {Popova},\ and\ \citenamefont {Lu}}]{gallmeier2006neutron}%
  \BibitemOpen
  \bibfield  {author} {\bibinfo {author} {\bibfnamefont {F.~X.}\ \bibnamefont
  {Gallmeier}}, \bibinfo {author} {\bibfnamefont {P.~D.}\ \bibnamefont
  {Ferguson}}, \bibinfo {author} {\bibfnamefont {E.~B.}\ \bibnamefont
  {Iverson}}, \bibinfo {author} {\bibfnamefont {I.~I.}\ \bibnamefont {Popova}},
  \ and\ \bibinfo {author} {\bibfnamefont {W.}~\bibnamefont {Lu}},\ }\href@noop
  {} {\bibfield  {journal} {\bibinfo  {journal} {Nuclear Instruments and
  Methods in Physics Research Section A: Accelerators, Spectrometers, Detectors
  and Associated Equipment}\ }\textbf {\bibinfo {volume} {562}},\ \bibinfo
  {pages} {946} (\bibinfo {year} {2006})}\BibitemShut {NoStop}%
\bibitem [{\citenamefont {Stone}\ \emph {et~al.}(2019)\citenamefont {Stone},
  \citenamefont {Crow}, \citenamefont {Fanelli},\ and\ \citenamefont
  {Niedziela}}]{stone2019characterization}%
  \BibitemOpen
  \bibfield  {author} {\bibinfo {author} {\bibfnamefont {M.~B.}\ \bibnamefont
  {Stone}}, \bibinfo {author} {\bibfnamefont {L.}~\bibnamefont {Crow}},
  \bibinfo {author} {\bibfnamefont {V.~R.}\ \bibnamefont {Fanelli}}, \ and\
  \bibinfo {author} {\bibfnamefont {J.~L.}\ \bibnamefont {Niedziela}},\
  }\href@noop {} {\bibfield  {journal} {\bibinfo  {journal} {Nuclear
  Instruments and Methods in Physics Research Section A: Accelerators,
  Spectrometers, Detectors and Associated Equipment}\ }\textbf {\bibinfo
  {volume} {946}},\ \bibinfo {pages} {162708} (\bibinfo {year}
  {2019})}\BibitemShut {NoStop}%
\bibitem [{\citenamefont {Stone}\ \emph {et~al.}(2015)\citenamefont {Stone},
  \citenamefont {Niedziela}, \citenamefont {Overbay},\ and\ \citenamefont
  {Abernathy}}]{stone2015arcs}%
  \BibitemOpen
  \bibfield  {author} {\bibinfo {author} {\bibfnamefont {M.~B.}\ \bibnamefont
  {Stone}}, \bibinfo {author} {\bibfnamefont {J.~L.}\ \bibnamefont
  {Niedziela}}, \bibinfo {author} {\bibfnamefont {M.~A.}\ \bibnamefont
  {Overbay}}, \ and\ \bibinfo {author} {\bibfnamefont {D.~L.}\ \bibnamefont
  {Abernathy}},\ }in\ \href@noop {} {\emph {\bibinfo {booktitle} {EPJ Web of
  Conferences}}},\ Vol.~\bibinfo {volume} {83}\ (\bibinfo {organization} {EDP
  Sciences},\ \bibinfo {year} {2015})\ p.\ \bibinfo {pages} {03014}\BibitemShut
  {NoStop}%
\bibitem [{\citenamefont {Haberl}\ \emph {et~al.}(2021)\citenamefont {Haberl},
  \citenamefont {Molaison}, \citenamefont {Frontzek}, \citenamefont {Novak},
  \citenamefont {Granroth}, \citenamefont {Goldsby}, \citenamefont {Anderson},\
  and\ \citenamefont {Elliott}}]{haberl20213d}%
  \BibitemOpen
  \bibfield  {author} {\bibinfo {author} {\bibfnamefont {B.}~\bibnamefont
  {Haberl}}, \bibinfo {author} {\bibfnamefont {J.~J.}\ \bibnamefont
  {Molaison}}, \bibinfo {author} {\bibfnamefont {M.}~\bibnamefont {Frontzek}},
  \bibinfo {author} {\bibfnamefont {E.~C.}\ \bibnamefont {Novak}}, \bibinfo
  {author} {\bibfnamefont {G.~E.}\ \bibnamefont {Granroth}}, \bibinfo {author}
  {\bibfnamefont {D.}~\bibnamefont {Goldsby}}, \bibinfo {author} {\bibfnamefont
  {D.~C.}\ \bibnamefont {Anderson}}, \ and\ \bibinfo {author} {\bibfnamefont
  {A.~M.}\ \bibnamefont {Elliott}},\ }\href@noop {} {\bibfield  {journal}
  {\bibinfo  {journal} {Review of Scientific Instruments}\ }\textbf {\bibinfo
  {volume} {92}} (\bibinfo {year} {2021})}\BibitemShut {NoStop}%
\bibitem [{\citenamefont {Stone}\ \emph {et~al.}(2014)\citenamefont {Stone},
  \citenamefont {Niedziela}, \citenamefont {Loguillo}, \citenamefont
  {Overbay},\ and\ \citenamefont {Abernathy}}]{stone2014radial}%
  \BibitemOpen
  \bibfield  {author} {\bibinfo {author} {\bibfnamefont {M.~B.}\ \bibnamefont
  {Stone}}, \bibinfo {author} {\bibfnamefont {J.~L.}\ \bibnamefont
  {Niedziela}}, \bibinfo {author} {\bibfnamefont {M.~J.}\ \bibnamefont
  {Loguillo}}, \bibinfo {author} {\bibfnamefont {M.~A.}\ \bibnamefont
  {Overbay}}, \ and\ \bibinfo {author} {\bibfnamefont {D.~L.}\ \bibnamefont
  {Abernathy}},\ }\href@noop {} {\bibfield  {journal} {\bibinfo  {journal}
  {Review of Scientific Instruments}\ }\textbf {\bibinfo {volume} {85}},\
  \bibinfo {pages} {085101} (\bibinfo {year} {2014})}\BibitemShut {NoStop}%
\bibitem [{\citenamefont {Copley}\ and\ \citenamefont
  {Cook}(1994)}]{copley1994analysis}%
  \BibitemOpen
  \bibfield  {author} {\bibinfo {author} {\bibfnamefont {J.}~\bibnamefont
  {Copley}}\ and\ \bibinfo {author} {\bibfnamefont {J.}~\bibnamefont {Cook}},\
  }\href@noop {} {\bibfield  {journal} {\bibinfo  {journal} {Nuclear
  Instruments and Methods in Physics Research Section A: Accelerators,
  Spectrometers, Detectors and Associated Equipment}\ }\textbf {\bibinfo
  {volume} {345}},\ \bibinfo {pages} {313} (\bibinfo {year}
  {1994})}\BibitemShut {NoStop}%
\bibitem [{SEs()}]{SEs}%
  \BibitemOpen
  \href@noop {} {}\bibinfo {howpublished}
  {https://neutrons.ornl.gov/sample/subpage/low-temperature-magnetic-fields)},\
  \bibinfo {note} {accessed: 2025-2-01}\BibitemShut {NoStop}%
\bibitem [{\citenamefont {Nakamura}\ \emph {et~al.}(2018)\citenamefont
  {Nakamura}, \citenamefont {Kambara}, \citenamefont {Iida}, \citenamefont
  {Kajimoto}, \citenamefont {Kamazawa}, \citenamefont {Ikeuchi}, \citenamefont
  {Ishikado},\ and\ \citenamefont {Aoyama}}]{nakamura2018performances}%
  \BibitemOpen
  \bibfield  {author} {\bibinfo {author} {\bibfnamefont {M.}~\bibnamefont
  {Nakamura}}, \bibinfo {author} {\bibfnamefont {W.}~\bibnamefont {Kambara}},
  \bibinfo {author} {\bibfnamefont {K.}~\bibnamefont {Iida}}, \bibinfo {author}
  {\bibfnamefont {R.}~\bibnamefont {Kajimoto}}, \bibinfo {author}
  {\bibfnamefont {K.}~\bibnamefont {Kamazawa}}, \bibinfo {author}
  {\bibfnamefont {K.}~\bibnamefont {Ikeuchi}}, \bibinfo {author} {\bibfnamefont
  {M.}~\bibnamefont {Ishikado}}, \ and\ \bibinfo {author} {\bibfnamefont
  {K.}~\bibnamefont {Aoyama}},\ }\href@noop {} {\bibfield  {journal} {\bibinfo
  {journal} {Physica B: Condensed Matter}\ }\textbf {\bibinfo {volume} {551}},\
  \bibinfo {pages} {480} (\bibinfo {year} {2018})}\BibitemShut {NoStop}%
\bibitem [{\citenamefont {Issa}\ \emph {et~al.}(2017)\citenamefont {Issa},
  \citenamefont {Khaplanov}, \citenamefont {Hall-Wilton}, \citenamefont
  {Llamas}, \citenamefont {Riktor}, \citenamefont {Brattheim},\ and\
  \citenamefont {Perrey}}]{issa2017characterization}%
  \BibitemOpen
  \bibfield  {author} {\bibinfo {author} {\bibfnamefont {F.}~\bibnamefont
  {Issa}}, \bibinfo {author} {\bibfnamefont {A.}~\bibnamefont {Khaplanov}},
  \bibinfo {author} {\bibfnamefont {R.}~\bibnamefont {Hall-Wilton}}, \bibinfo
  {author} {\bibfnamefont {I.}~\bibnamefont {Llamas}}, \bibinfo {author}
  {\bibfnamefont {M.~D.}\ \bibnamefont {Riktor}}, \bibinfo {author}
  {\bibfnamefont {S.}~\bibnamefont {Brattheim}}, \ and\ \bibinfo {author}
  {\bibfnamefont {H.}~\bibnamefont {Perrey}},\ }\href@noop {} {\bibfield
  {journal} {\bibinfo  {journal} {Physical Review Accelerators and Beams}\
  }\textbf {\bibinfo {volume} {20}},\ \bibinfo {pages} {092801} (\bibinfo
  {year} {2017})}\BibitemShut {NoStop}%
\bibitem [{\citenamefont {Nakamura}\ \emph {et~al.}(2015)\citenamefont
  {Nakamura}, \citenamefont {Kawakita}, \citenamefont {Kambara}, \citenamefont
  {Aoyama}, \citenamefont {Kajimoto}, \citenamefont {Nakajima}, \citenamefont
  {Ohira-Kawamura}, \citenamefont {Ikeuchi}, \citenamefont {Kikuchi},
  \citenamefont {Inamura} \emph {et~al.}}]{nakamura2015oscillating}%
  \BibitemOpen
  \bibfield  {author} {\bibinfo {author} {\bibfnamefont {M.}~\bibnamefont
  {Nakamura}}, \bibinfo {author} {\bibfnamefont {Y.}~\bibnamefont {Kawakita}},
  \bibinfo {author} {\bibfnamefont {W.}~\bibnamefont {Kambara}}, \bibinfo
  {author} {\bibfnamefont {K.}~\bibnamefont {Aoyama}}, \bibinfo {author}
  {\bibfnamefont {R.}~\bibnamefont {Kajimoto}}, \bibinfo {author}
  {\bibfnamefont {K.}~\bibnamefont {Nakajima}}, \bibinfo {author}
  {\bibfnamefont {S.}~\bibnamefont {Ohira-Kawamura}}, \bibinfo {author}
  {\bibfnamefont {K.}~\bibnamefont {Ikeuchi}}, \bibinfo {author} {\bibfnamefont
  {T.}~\bibnamefont {Kikuchi}}, \bibinfo {author} {\bibfnamefont
  {Y.}~\bibnamefont {Inamura}},  \emph {et~al.},\ }in\ \href@noop {} {\emph
  {\bibinfo {booktitle} {Proceedings of the 2nd International Symposium on
  Science at J-PARC-Unlocking the Mysteries of Life, Matter and the
  Universe}}}\ (\bibinfo {year} {2015})\ p.\ \bibinfo {pages}
  {36011}\BibitemShut {NoStop}%
\bibitem [{\citenamefont {Qiu}\ \emph {et~al.}(2022)\citenamefont {Qiu},
  \citenamefont {Hu}, \citenamefont {Liu}, \citenamefont {Zhou}, \citenamefont
  {Shen}, \citenamefont {Jin}, \citenamefont {Du}, \citenamefont {Ding},
  \citenamefont {Song}, \citenamefont {Huang} \emph {et~al.}}]{qiu2022neutron}%
  \BibitemOpen
  \bibfield  {author} {\bibinfo {author} {\bibfnamefont {J.}~\bibnamefont
  {Qiu}}, \bibinfo {author} {\bibfnamefont {C.}~\bibnamefont {Hu}}, \bibinfo
  {author} {\bibfnamefont {Y.}~\bibnamefont {Liu}}, \bibinfo {author}
  {\bibfnamefont {L.}~\bibnamefont {Zhou}}, \bibinfo {author} {\bibfnamefont
  {F.}~\bibnamefont {Shen}}, \bibinfo {author} {\bibfnamefont {D.}~\bibnamefont
  {Jin}}, \bibinfo {author} {\bibfnamefont {W.}~\bibnamefont {Du}}, \bibinfo
  {author} {\bibfnamefont {C.}~\bibnamefont {Ding}}, \bibinfo {author}
  {\bibfnamefont {W.}~\bibnamefont {Song}}, \bibinfo {author} {\bibfnamefont
  {Y.}~\bibnamefont {Huang}},  \emph {et~al.},\ }\href@noop {} {\bibfield
  {journal} {\bibinfo  {journal} {Nuclear Instruments and Methods in Physics
  Research Section A: Accelerators, Spectrometers, Detectors and Associated
  Equipment}\ }\textbf {\bibinfo {volume} {1033}},\ \bibinfo {pages} {166684}
  (\bibinfo {year} {2022})}\BibitemShut {NoStop}%
\end{thebibliography}
\end{document}